\documentclass[
    aps,                
    pra,                
    twocolumn,          
    superscriptaddress, 
    floatfix            
    11pt,
    notitlepage,
    tightenlines,
    nofootinbib,
    reprint
]{revtex4-2}

\usepackage{natbib}

\usepackage{comment}

\usepackage[
    pdftex
]{hyperref}
\hypersetup{
    colorlinks,
    linkcolor=[rgb]{0.206756, 0.371758, 0.553117},
    urlcolor=[rgb]{0.120565, 0.596422, 0.543611},
    citecolor=[rgb]{0.8784313725490196, 0.403921568627451, 0.4549019607843137},
    breaklinks=true 
}

\usepackage{amsfonts, amssymb, amsmath, amsthm}
\usepackage{mathrsfs}               
\usepackage{bbold}
\usepackage{bm}                     
\usepackage{physics}
\usepackage{dsfont}                 
\usepackage[version=4]{mhchem}      
\usepackage[normalem]{ulem}
\usepackage[a4paper]{geometry}
\usepackage{graphicx}               
\usepackage{svg}
\usepackage{caption}
\usepackage{subcaption}
\usepackage{ragged2e} 
\usepackage{float}
\usepackage{color}
\usepackage{xcolor}

\usepackage{multirow}
\usepackage{array}
\usepackage{booktabs}
\usepackage{tabularx}
\usepackage{adjustbox}
\newcolumntype{L}[1]{>{\raggedright\let\newline\\\arraybackslash\hspace{0pt}}m{#1}}
\newcolumntype{C}[1]{>{\centering\let\newline\\\arraybackslash\hspace{0pt}}m{#1}}
\newcolumntype{R}[1]{>{\raggedleft\let\newline\\\arraybackslash\hspace{0pt}}m{#1}}

\DeclareEmphSequence{\bfseries\itshape}

\newcommand{\figref}[1]{Fig.~\ref{#1}}    
\newcommand{\Eqref}[1]{Eq.~\eqref{#1}}    
\newcommand{\tabref}[1]{Tab.~\ref{#1}}    

\newcommand{\veps}{\varepsilon}
\newcommand{\vphi}{\varphi}
\renewcommand{\dd}{\mathrm{d}}        
\newcommand{\identity}{\mathds{1}}
\newcommand{\mean}[1]{\left<#1\right>}

\newcommand{\q}[1]{``#1''}          
\renewcommand{\abs}[1]{\left|#1\right|}
\renewcommand{\norm}[1]{\left\lVert#1\right\rVert}
\renewcommand{\op}[1]{\hat{#1}}

\newcommand{\si}[1]{\, \mathrm{#1}} 
\newcommand{\Si}[1]{\mathrm{#1}}    

\DeclareMathOperator{\diag}{diag}

\DeclareMathOperator{\Cov}{Cov}

\theoremstyle{plain}

\theoremstyle{definition}

\theoremstyle{remark}

\begin{document}
\newcommand{\ITP}{Institut f\"{u}r Theoretische Physik, Universit\"{a}t Ulm, Albert-Einstein-Allee 11, D-89069 Ulm, Germany}
\newcommand{\IQST}{Center for Integrated Quantum Science and Technology (IQST), 89081 Ulm, Germany}

\title{Stability Thresholds for Gravitationally Induced Entanglement in Shielded Setups}

\author{Jan Bulling}
\email{jan.bulling@uni-ulm.de}

\author{Marit O. E. Steiner}
\email{marit.steiner@uni-ulm.de}

\author{Julen S. Pedernales}
\email{julen.pedernales@uni-ulm.de}

\author{Martin B. Plenio}
\email{martin.plenio@uni-ulm.de}
\affiliation{\ITP}
\affiliation{\IQST}

\date{\today}

\begin{abstract}
    Proposed experiments for gravitationally induced entanglement (GIE) typically suppress direct electromagnetic interactions between two massive particles by inserting a conducting Faraday shield. For superconducting particles, their large diamagnetism requires additional magnetic shielding to screen magnetic dipolar interactions.
    Here, we analyze the effect of residual particle--shield interactions and show that both Casimir and magnetic-dipole interactions can severely limit GIE tests by imprinting large phases. 
    We quantify how run-to-run positional and orientational fluctuations of the setup elements, including the shield, trapping potentials, and detectors, convert these phases into effective decoherence, strongly reducing the detectable bipartite entanglement.
    In particular, we show that magnetic interactions between the particles and a superconducting shield constitute a major noise source, especially relevant for levitated superconducting particles.
    Treating the vibrational modes of the shield quantum mechanically, we further find that thermal vibrations generate persistent particle--shield correlations and can even mediate particle--particle entanglement that can mimic a gravitational signal.
    Finally, we derive quantitative thresholds on the maximum tolerable positional and orientational fluctuations of the setup elements required to observe entanglement, and propose mitigation strategies including geometry optimization and shield cooling to preserve a genuine GIE signature.
\end{abstract}

\maketitle

\section{Introduction}\label{sec:introduction}
Despite the success of quantum theory in describing the electromagnetic, weak, and strong interactions, a complete quantum theory of gravity valid at all energy scales remains elusive. In experimentally tested regimes, gravity is well described by classical general relativity.
Exploring physical regimes beyond those tested to date, in which both quantum mechanics and gravity become simultaneously relevant, may enable tests of the quantum nature of gravity.
A promising low-energy route toward such tests was originally proposed by Feynman as a gedankenexperiment nearly seventy years ago~\cite{DeWitt2017} and later explicitly connected to gravitational entanglement generation in the quantum information community~\cite{Lindner2005,Kafri2013}.
It aims to observe the generation of entanglement between two massive particles, each prepared with its center of mass coherently delocalized in position, and interacting solely through gravity.
The observation of such gravitationally induced entanglement (GIE) between the particles’ positional degrees of freedom would imply that gravitational interaction cannot be described by local operations and classical communication (LOCC) alone, and would therefore exhibit inherently non-classical features~\cite{Christodoulou2023, PlenioV2007}.

In light of recent advances in quantum control of levitated massive systems~\cite{Delic2020, GonzalezBallestero2021, Kamba2023, Piotrowski2023, Ranfagni2022, Dania2025, Magrini2021, Tebbenjohanns2021,  Kamba2025, Troyer2026} and in precision gravity measurements at small scales~\cite{Westphal2021, Lee2020}, such tests have attracted growing theoretical and experimental interest, despite the daunting practical challenges. A variety of experimental proposals have been put forward across different platforms~\cite{Lindner2005, Pino2018, Lami2024, Li2025, Howl2019, Howl2021, Oppenheim2023}, often involving micrometer-sized nanoparticles with masses in the pico- to nanogram range prepared in spatially delocalized quantum states~\cite{Bose2017, Pedernales2022, Pedernales2023, Lantano2024, Kaltenbaek2012, Shiomatsu2025, Petruzziello2025}.
The central intuition behind many of these schemes is that, even in the weak-field Newtonian regime, quantum gravity provides an effective interaction that can generate entanglement~\cite{Christodoulou2023, Carney2019, Danielson2022} (see Refs.~\cite{Lantano2024,Petruzziello2025} for extensions to general relativistic effects).
The present work focuses on implementations based on levitated nanoparticles, which combine excellent isolation from the environment with precise control over their motional quantum states -- both essential for detecting the extremely weak gravitational coupling between the particles.
We analyze the degree of control required to successfully perform such experiments.
While our analysis is tailored to this platform, the results are in principle transferable to a broader class of experimental realizations, including free-fall and space-based implementations.

A notable challenge in such experiments arises from short-range electromagnetic interactions.
At micrometer-scale separations, electrostatic, magnetostatic and Casimir forces can easily dominate over the gravitational coupling.
These unwanted interactions can imprint additional dynamical phases or even generate unwanted entanglement, thereby obscuring the gravitational signal.
To overcome this, experimental proposals typically include a thin conducting or superconducting shield between the particles, suppressing electromagnetic cross-talk while leaving gravity unaffected~\cite{Schmoele2016, Kamp2020, Westphal2021}.

Introducing such a shield, however, also brings its own complications.
Aside of material specific decoherence mechanisms and patch potentials, which we neglect in our study, Casimir and magnetic-dipole forces between each particle and the shield generate a static force that shifts the particle’s center-of-mass position and induce a local phase that depends on the particle--shield distance.
For a spatially delocalized particle, different components of the wavefunction experience different particle--shield separations, resulting in a relative phase accumulation across the superposition that is imprinted locally on each particle.
While the displacement due to the static interaction can be suppressed by sufficiently stiff trapping potentials, the phase sensitivity becomes problematic in realistic experiments: slight run-to-run fluctuations in the geometric configuration of the setup lead to variations of these induced phases.
When averaged over many repetitions, these fluctuations manifest as dephasing, thereby degrading the experimentally verifiable gravitationally induced entanglement.

In this work, we provide a quantitative analysis of these effects:
We consider stochastic forces acting on the setup and estimate the impact of fluctuations in the relative positioning of the trapping potentials, the shield and the measurement devices on the quantum coherence of the system.
Both, Casimir and magnetic dipole interactions between the particles and the shield are taken into account.
Additionally, we consider the quantized vibrational modes of the shield and analyze noise arising from interactions with them, as well as shield-mediated, non-gravitational entanglement.

The article is organized as follows.
In Sec.~\ref{sec:setup}, we describe the class of setups considered and analyze the interactions that can arise between the particles and the shield, as well as between the particles themselves, discussing their origin and estimating their strength.
In Sec.~\ref{sec:variations}, we examine run-to-run positional and orientational fluctuations of the setup elements and quantify how, in the presence of the interactions identified in Sec.~\ref{sec:setup}, such fluctuations suppress the achievable gravitationally induced entanglement.
We carry out this analysis for both squeezed Gaussian states and Schrödinger-cat states, deriving quantitative bounds on the tolerable fluctuations.
In Sec.~\ref{sec:trap-stability}, we turn to fluctuations of the trapping potentials, which induce noise even in the absence of particle--shield interactions, and determine the corresponding stability thresholds.
In Sec.~\ref{sec:shield}, we consider a dynamical shield with quantum mechanical degrees of freedom, allowing it to become correlated with the particles.
We quantify the resulting noise as a function of the shield’s temperature and geometry, and analyze the non-gravitational quantum channel it mediates, bounding the entanglement it can generate.
Finally, in Sec.~\ref{sec:implications}, we summarize the experimental implications of our results, and we conclude in Sec.~\ref{sec:conclusion}.
\section{The Setup}\label{sec:setup}
\begin{figure*}[!th]
    \centering
    \includegraphics[width=\linewidth]{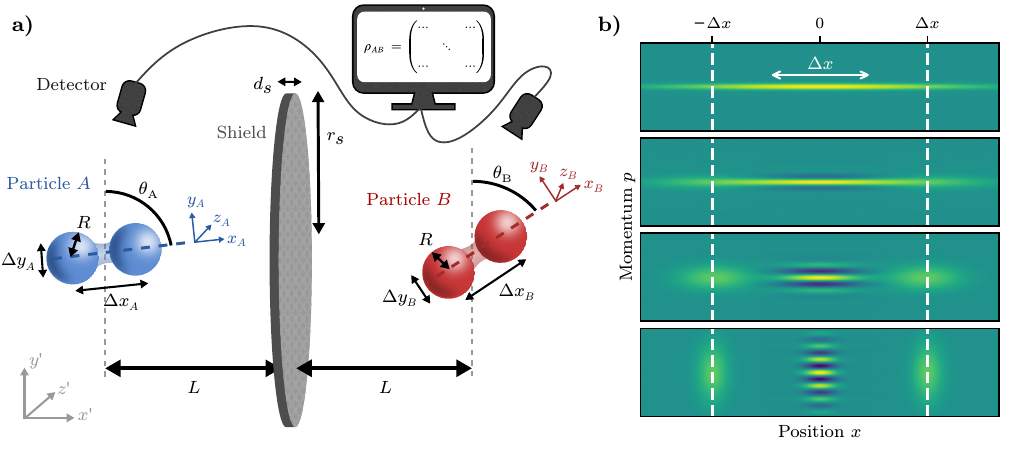}
    \caption{\justifying\small
    \textbf{Setup and initial states.}
    \textbf{a,} Schematic illustration of an idealized setup for gravitationally induced entanglement (GIE) experiments considered in this work.
    Two particles $A$ and $B$ of radius $R$ are prepared in spatially delocalized quantum states and positioned at a fixed center-to-center separation $2L$. 
    A circular shield of thickness $d_s$ and radius $r_s$ is placed midway between the particles to suppress unwanted electromagnetic crosstalk while preserving the gravitational coupling.
    The angles $\theta_{A/B}$ indicate the orientation of each particle’s principal delocalization axis (here without loss of generality the $x$-direction) and the shield.
    Two detectors measure the state of the particles after a gravitational interaction.
    \textbf{b,} Wigner phase-space representation of the family of delocalized states with fixed  standard deviation $\Delta x$ used in this work, ranging from squeezed Gaussian states \textbf{(top)} to symmetric superpositions of two displaced squeezed states \textbf{(bottom)}, illustrating the continuous interpolation up to idealized orthonormal cat-states representable in a truncated 2-dimensional Hilbert space.}
    \label{fig:setup}
\end{figure*}

In this analysis, we consider an idealized model of the GIE experiment, abstracting away implementation-specific details.
The schematics are presented in Fig.~\ref{fig:setup}(a).
The setup consists of two spherical nanoparticles, labeled $A$ and $B$, which are separated by a fixed center-to-center distance $2L$. 
We introduce a local coordinate system $(x,y,z)_{A/B}$ for each particle and denote the laboratory frame by the primed coordinates $(x',y',z')$. After letting the particles interact gravitationally, their state is read-out by two detectors.
In order to suppress non-gravitational interactions, a circular shield of radius $r_s$ and thickness $d_s$ is placed symmetrically midway between the particles, with its surface lying in the $y'z'$-plane.

We consider both, silica and superconducting lead nanoparticles with a radius of $R=10\si{\mu m}$, corresponding to masses in the range $M\approx 1\times10^{-11}\si{kg}$ to $5\times 10^{-11}\si{kg}$. 
Superconducting lead particles, which are strongly diamagnetic and typically magnetically levitated, serve as a representative example of regimes in which particle--shield interactions are dominated by induced magnetic dipole moments.
In contrast, in the absence of magnetic trapping, e.g., in free-falling or space-based implementations, weakly diamagnetic silica particles serve as a representative example of regimes in which particle--shield interactions are dominated by Casimir forces.
\begin{table*}[!th]
    \caption{\justifying\small
    \textbf{Set of parameters}
    We consider silica nanoparticles ($\rho_\mathrm{SiO_2}=2650\si{kg/m^3}$) in free-fall or space-based implementations and superconducting lead particles ($\rho_\mathrm{Pb}=11340\si{kg/m^3}$) levitated magnetically (with a static trapping field $B_\mathrm{trap} = 250\si{\mu T}$).
    Both particles are chosen to have a radius of $R=10\si{\mu m}$ for better comparability, as the relevant interactions \Eqref{eq:casimir-pfa} and \Eqref{eq:mag-dipole-potential} depend on the radius.
    In principle, Silica particles could also be magnetically levitated with a static trapping field or $B_\mathrm{trap}\simeq 20\si{mT}$.
    }
    \label{tab:parameters}
    \begin{ruledtabular}
    \begin{tabular}{c l c c c c c c c c}
        \multicolumn{5}{c}{Particle properties} & \multirow{2}{*}{$L$} & \multirow{2}{*}{$\Delta x$} & \multirow{2}{*}{$\Delta y$} & \multicolumn{2}{c}{Shield properties} \\ \cline{1-5} \cline{9-10}
         Material & Dominant interaction & Mass $M$ / $\Si{kg}$ & $\veps_r$ & $\chi_V$ & & & & Radius $r_s$ & Thickness $d_s$ \\ \hline
         Silica ($\text{SiO}_2$) & Casimir forces & $1.11\times 10^{-11}$ & $3.9$ & $-1.4\times 10^{-5}$ & \multirow{2}{*}{$20\si{\mu m}$} & \multirow{2}{*}{$50\si{nm}$} & \multirow{2}{*}{$10^{-12}\si{m}$} & \multirow{2}{*}{$1\si{cm}$} & \multirow{2}{*}{$2\si{\mu m}$} \\
         Lead (Pb) & induced mag. dipoles & $4.75\times 10^{-11}$ & $\infty$ & -1 & & & & &
    \end{tabular}
    \end{ruledtabular}
\end{table*}

The particles are prepared in spatially delocalized pure quantum states of their center-of-mass motion.
We consider squeezed Gaussian states and Schrödinger-cat-like superpositions, characterized by position uncertainties $\Delta x_{A/B}$ and $\Delta y_{A/B}$ defined along the axes of each particle’s local coordinate frame.
Without loss of generality, we take $\Delta x \gg \Delta y$, thereby identifying the local $x$ axis as the principal delocalization direction.
In the lab frame, the orientation of the initial state is parameterized by a single angle $\theta_{A/B} \in [0,\pi)$, corresponding to rotations of the state about the $z$-axis.
The delocalization direction is thus confined to the $x'y'$-plane, while any $z'$-component is neglected.
This restriction is well justified in the regime $\Delta z \lesssim \Delta y \ll \Delta x$.
Moreover, displacements along the $z$-axis parallel to the shield do not significantly affect the relevant geometries, provided that $\Delta z \ll r_s$ and the particles remain sufficiently far from the shield boundaries.
A fully general treatment with arbitrary spatial spreads and orientations would require the inclusion of all three Euler angles.
However, such an extension does not provide additional physical insight in the regime of interest and is therefore not pursued.

For electrostatic and Casimir screening, a conducting Faraday shield is used, which may be fabricated from copper~\cite{Kamp2020}, gold-coated substrates~\cite{Schut2023} or graphene~\cite{Pavlou2021}. On the other hand, magnetic interactions require superconducting Meissner shielding, for example based on superconducting type-II Niobium.

Since all relevant interactions differ fundamentally in range and strength, the required shield dimensions depend on the dominant interaction, as discussed in detail in the subsequent sections.
In practice, however, the minimum shield thickness is set by mechanical rigidity and fabrication constraints, which impose a larger thickness than that required by the shielding performance.
In the following, we assume a thickness of $d_s = 2\si{\mu m}$, which is sufficient to shield all undesired electromagnetic interactions.

We focus on two geometries commonly discussed in GIE proposals:
In the \textit{parallel configuration} ($\theta_{A/B}=0$), the principal delocalization axes of the two particles are parallel to each other and to the shield surface.
In the \textit{linear orientation} ($\theta_{A/B}=\pi/2$), the delocalization axes of both particles coincide, forming a single line perpendicular to the shield.
As discussed later, the choice between these configurations has important consequences for the robustness of entanglement generation in presence of shield-induced noise.

\subsection{Casimir interactions}
At surface-to-surface separations below a few hundred micrometers, short-range Casimir forces~\cite{Casimir1948} constitute an important source of non-gravitational interactions between the nanoparticles.
Neglecting finite-frequency penetration of electromagnetic fields into the material, mutual particle--particle interactions can be efficiently suppressed by inserting a conductive Faraday shield of radius $r_s\gtrsim 50\si{\mu m}$ and thickness $d_s\gtrsim 3\times 10^{-10}\si{m}$, as estimated in Appendix~\ref{apx:shield-size} and summarized in \tabref{tab:shield-size}.
In practice, however, the finite optical skin depth imposes a more stringent lower bound on the thickness of around $d_s\approx 20\si{nm}$~\cite{Roberts1960}.

Casimir interactions between each particle and the shield must nevertheless be taken into account.
Closed-form expressions for this potential in the sphere-plane geometry are not available for arbitrary separations.
Analytic results exist only in the limiting cases of short and large separations, corresponding to the proximity-force approximation (PFA) valid for small surface-to-surface separations $L - R \ll R$, and the large-separation limit (LSL), valid in the opposite regime $L \gg R$.
In this work, we operate in an intermediate regime where neither limit is valid.
We therefore employ the proximity-force approximation, which overestimates the actual Casimir strength for the parameters in \tabref{tab:parameters} by a factor of $\approx2.4$ and thus provides a conservative upper bound on decoherence effects, as discussed in Appendix~\ref{apx:interactions-casimir}.
In the sphere-plane geometry, it is given by~\cite{Bulgac2006, Pirozhenko2013}
\begin{align}\label{eq:casimir-pfa}
    V_\mathrm{PFA} = -\frac{\hbar c \pi^3 R}{720 (L - R - d_s/2)^2} \, \varphi(\varepsilon_r)
\end{align}
where $L-R-d_s/2$ is the surface-to-surface separation of the sphere and the plane.
The factor $\varphi(\varepsilon_r)$ accounts for dielectric properties of the particles as a function of the relative permittivity $\varepsilon_r$ and is given by $\varphi(\varepsilon_r)=(\varepsilon_r-1)/(\varepsilon_r+1)\chi(\varepsilon_r)$ with $\chi(\varepsilon_r)$ being a tabulated function~\cite{Lifshitz1992}.
In particular, $\chi(\varepsilon_r \rightarrow\infty)\rightarrow 1$, $\chi(\varepsilon_r = 1) \approx 0.46$ and $\chi(\varepsilon_{r,\,\text{SiO}_2})\approx 0.5$.
Finite temperature corrections to the Casimir interaction exists but are negligible for temperatures $T \ll \hbar c / (k_B L)\approx 100\si{K}$~\cite{Bordag2009}.

\subsection{Magnetic interactions}
In addition to Casimir forces, induced magnetic-dipole moments arising from external magnetic fields, e.g. the magnetic trapping potential, introduce an additional source of non-gravitational coupling between the particles in the form of magnetic dipole-dipole interactions. These can quickly become dominant over the gravitational interaction and thus also need to be suppressed.
Magnetic interactions are however only weakly shielded by a conducting Faraday shield and therefore require explicit consideration.
Furthermore, the relevance of these interactions depends crucially on the particle material and differ quantitatively between the silica particles and the superconducting lead particles considered in this work.
For a spherical particle of radius $R$ and volume magnetic susceptibility $\chi_V$, an external magnetic field $\vec{B}$ induces a magnetic-dipole moment $\vec{m}=4\pi R^3\chi_V\vec{B}/(3\mu_0)$, where $\mu_0$ is the vacuum permeability.
Silica is weakly diamagnetic with $\chi_{V,\,\text{SiO}_2} \approx -1.4\times 10^{-5}$~\cite{CRC2017}, whereas superconducting lead behaves as a perfect diamagnet with $\chi_V = -1$.

Requiring the magnetic dipole-dipole interaction to remain smaller than gravitational coupling imposes an upper bound on the admissible external magnetic field.
As shown in Appendix~\ref{apx:shield-size}, this condition yields $B \lesssim 71\si{\mu T}$ for silica particles, which is on par with Earth's magnetic field.
For superconducting lead particles, the bound is significantly more stringent with $B\lesssim 4.25 \si{nT}$.
The screening of magnetic dipole-dipole interactions between the particles and of external magnetic fields is therefore necessary.

Besides the Earth's magnetic field, additional fields may arise from the trapping mechanism itself.
For diamagnetically levitated particles, static magnetic fields are required to levitate, following
\begin{equation*}
    B \partial_x B = \frac{g \rho \mu_0}{\chi_{V}},
\end{equation*}
where $g=9.81\si{m/s^2}$.
To levitate superconducting lead with magnetic gradients of $\partial_x B \sim 10^2-10^3\si{T/m}$~\cite{Naito2024} requires a static magnetic field of $B_\mathrm{trap}\sim 0.1 - 1\,\mathrm{m T}$.
For the less diamagnetic silica, however, higher gradients of $\partial_x B \sim 10^4-10^5\si{T/m}$~\cite{Hsu2016} with static magnetic fields of $B_\mathrm{trap}\sim 20-200\si{mT}$ are sufficient
\footnote{In the absence of gravity, a diamagnetic particle can be trapped at $B=0$, for example at the center of an ideal anti-Helmholtz configuration.
In this case, the induced response is purely quadrupolar, leading to comparatively weaker interactions with the shield.
In terrestrial experiments, however, gravity is unavoidable and shifts the equilibrium position away from the field zero, thereby inducing a finite dipolar contribution.}.

When the magnetic moments are aligned, the dipole-dipole interaction is proportional to $3\cos^2\phi - 1$, where $\phi$ is the angle between the dipole orientation and the interparticle separation vector.
It vanishes at $\phi \approx 54.7^\circ$, the \q{magic angle}.
However, achieving this configuration would require precise control of the applied field direction, particularly for the superconducting lead particle, as well as of the experimental geometry.
Moreover, such cancellation is impossible for dipole components induced by uncontrolled stray fields.

An alternative approach is to modulate the induced dipoles in time, for instance via modulation of the trapping fields, such that the interaction averages to zero over a cycle.
While this dynamical averaging suppresses the controlled dipole-dipole coupling, it again leaves contributions from uncontrolled stray fields unaffected. We therefore do not pursue this mechanism further and instead focus on suppression via magnetic shielding.

One option is using high-permeability materials, such as Mu-metals with a permeability of $\mu_r\sim10^5$~\cite{CRC2017,Jiles2015}.
While such materials may suffice to screen weak magnetic interactions, they are generally ineffective in shielding strong couplings, like for magnetically levitated particles, where we assume a static trapping magnetic field of $B_\mathrm{trap} = 250\si{\mu T}$ is required.
For such regimes, Meissner shielding using a type-II superconductor is required.
Niobium ($T_c\approx 10\si{K}$) with a London-penetration depth of $\lambda_L \approx 40\si{nm}$~\cite{Maxfield1965} would sufficiently suppress magnetic couplings at a thickness of $d_s\gtrsim 440\si{nm}$, as seen in Appendix~\ref{apx:shield-size}.

However, introducing a magnetic shield, whether based on high-permeability materials or superconductors, inevitably leads to additional particle--shield interactions, analogous to the Casimir forces discussed previously.
The interaction potential from a magnetic dipole and its mirror dipole in the superconducting shield is given by
\begin{equation}\label{eq:mag-dipole-potential}
    V_\mathrm{mag.\,Dipole} = -\frac{\abs{\vec{m}}^2\mu_0}{32 \pi (L-d_s/2)^3} \frac{\mu_r - 1}{\mu_r + 1} \left(1 + \cos^2\phi \right)
\end{equation}
for a shield with magnetic permeability $\mu_r$, where $\phi$ is the angle between the magnetic-dipole moment and the shield's normal.
For the parameters in \tabref{tab:parameters} and $B_\mathrm{trap}=250\si{\mu T}$, these interactions between magnetically trapped superconducting lead particles and the shield exceed the corresponding Casimir interactions by a factor of $10^4$ and the direct gravitational particle--particle coupling by $10^9$.
For silica magnetically levitated in $B_\mathrm{trap}\lesssim 20\si{mT}$, Casimir interactions with the shield still dominates over magnetic dipole interactions.

\subsection{Electrostatic interactions}
Electrostatic forces pose a further challenge for GIE experiments due to the large disparity between electromagnetic and gravitational interaction strengths.
If the particles carry a net charge $q$, direct Coulomb interactions would easily dominate over gravity, unless effectively shielded.
Even for a single elementary charge $q=e$ on each particle, suppressing Coulomb couplings to below the gravitational interaction would require a Faraday shield of radius $r_s\gtrsim66\si{cm}$ and thickness $d_s\gtrsim 2 \times 10^{-8}\si{m}$, as estimated in Appendix~\ref{apx:shield-size}.
Such a large shield would support low-frequency vibrational modes that are thermally populated even at low temperatures and couple to the particles, thereby introducing additional noise, as discussed in Sec.~\ref{sec:shield}. 
We therefore strongly advocate operating with electrically neutral particles.

However, even in the absence of a net charge, couplings can arise from permanent or induced electric dipoles.
The particles may possess intrinsic dipole moments due to lattice or surface inhomogeneities~\cite{Priel2022} on the order of $p\sim eR$ (e.g. $p\approx 10^{-2}e\si{cm}$ in micrometer silica nanoparticles~\cite{Rider2016}).
Moreover, stray electric fields $\vec{E}$ can induce dipole moments as $\vec{p}=4\pi\varepsilon_0 (\varepsilon_r - 1)/(\varepsilon_r + 2) R^3\vec{E}$.
Particularly relevant sources of stray fields are spatially varying electrostatic patch potentials on the shield's surface.
These arise from crystal defects, mechanical stress, or surface contamination and typically exhibit voltage variations $V_\mathrm{Patch}\sim 1 - 100\si{mV}$~\cite{Robertson2007,Garrett2020} over length scales $\ell_\mathrm{Patch}\sim 100\si{nm}$.
Such potentials generate electric fields of magnitude $E\sim V_\mathrm{Patch}\ell_\mathrm{Patch} / L^2 \approx 10^{-1}-10^{2}\si{V/m}$ which can induce dipole moments $p \approx 10^{-5}-10^{-3}e\si{cm}$ in the particles.
Shielding all these dipolar interactions between the two particles to below the gravitational coupling requires a more modest shield size with radius $r_s\gtrsim 1\si{mm}$ and thickness $d_s\gtrsim 5\times 10^{-11}\si{m}$, as estimated  in Appendix~\ref{apx:shield-size} and summarized in \tabref{tab:shield-size}. 
In practice, the optical skin depth of the shielding material as well as rigidity and fabrication methods give stricter lower bound on the thickness than the physical shielding.

Each dipole also interacts with the conducting shield, resulting in an interaction potential given by
\begin{equation*}
    V_\mathrm{elec.\,Dipole} = -\frac{\abs{\vec{p}}^2}{32\pi\varepsilon_0  (L-d_s/2)^3}\left(1 + \cos^2\phi\right) ,
\end{equation*}
where $\phi$ again denotes the angle between the dipole orientation and the shield's normal.
For the parameters listed in \tabref{tab:parameters}, the dipole-induced particle--shield interaction is weaker than the corresponding Casimir interaction by a factor ranging from $\sim10^{-6}$ to $\sim 10^{0}$ and weaker than magnetic dipole interactions by $\sim 10^{-10}-10^{-4}$ (depending on the magnitude of the induced electric dipole moment).
Thus, we only consider Casimir- and magnetic-dipole induced effects in the following.
That the decoherence induced by electric dipole interactions could be analyzed the same way as in Sec.~\ref{sec:variations} and Sec.~\ref{sec:shield}.

\subsection{Gravitational interactions}
The gravitational interaction between the two particles is described by a Newtonian potential, as we operate in the nonrelativistic, low-energy regime:
\begin{equation}\label{eq:general-gravitational-hamiltonian}
\op{H}_\mathrm{Gravity} = -\frac{G M_A M_B}{\big|\op{d}_{AB}\big|} ,
\end{equation}
where $\op{d}_{AB}=\sqrt{(\hat{x}'_A - \hat{x}'_B)^2 + (\hat{y}'_A - \hat{y}'_B)^2}$ denotes the canonically quantized distance operator between the centers of mass of the two particles.
Provided that the delocalization of each particle is small compared to their separation, i.e., $\Delta x, \Delta y \ll L$, the potential in \Eqref{eq:general-gravitational-hamiltonian} can be expanded up to second order in $\op{x}$ and $\op{y}$.
As shown in Appendix~\ref{apx:interactions-gravity}, this expansion yields bilinear coupling terms of the form $\propto x_A x_B$ and $\propto y_A y_B$, which constitute the origin of gravitationally induced entanglement in this model~\cite{Pedernales2023}.

Gravitational interactions between the particles and the shield are neglected in this work, as they are several orders of magnitude weaker than all other particle--shield interactions discussed before.

Beyond mutual interaction between the particles, sources of external gravitational fields, ranging from the Earth and nearby celestial bodies to moving massive objects in the laboratory and its surroundings, can influence the phase evolution of the delocalized state.
Since the spatial superpositions must be maintained in a fixed configuration with respect to the experimental screen, they generally acquire relative phases due to these external gravitational accelerations.

In particular, the dominant contributions originate from the Sun and the Moon, with effective accelerations of approximately $g_\mathrm{Sun} \approx 6 \times 10^{-4}\, g$ and $g_\mathrm{Moon} \approx 3.3 \times 10^{-6}\, g$, while contributions from other planets are typically at the level of $10^{-8}\, g$ or smaller.
In the worst-case configuration, this leads to a relative phase accumulation rate given by
\begin{equation*}
    \Gamma = \frac{2\Delta x M g_{\mathrm{Sun}}}{\hbar}
\end{equation*}
for a particle of mass $M$.
For the parameters in \tabref{tab:parameters}, we find $\Gamma \approx 10^{13} - 10^{14}\si{Hz}$.
Such large phase evolution rates impose stringent requirements on the timing precision of the experiment, as well as on the accurate knowledge of the relative orientation between the experimental setup and the Sun.

In principle, this contribution could be reduced by using a test mass to measure the total acceleration arising from the combined gravitational influence of the Earth, the Sun and other bodies and subsequently aligning the principal delocalization axis orthogonal to this total acceleration vector.
At the same time, the shield imposes an additional alignment constraint, as its normal vector must be aligned accordingly with the delocalization axis, depending on wether the parallel or linear orientation is prepared.
This introduces a nontrivial geometrical constraint on the experimental design, where the shield has to co-rotate with the changing total acceleration vector.

While the severity of this effect is reduced for smaller test masses, it remains non-negligible and must be accounted for.
More generally, it is worth noting that even in exceptionally quiet underground environments or close to the first Lagrange point, residual random accelerations are typically of the order of $10^{-8}\si{m/s^2/\sqrt{Hz}}$ at $1\si{Hz}$~\cite{Harms2010, Armano2024}, providing an additional background contribution that limits phase stability.
\section{Variations in the experimental geometry}\label{sec:variations}
\begin{figure*}[!th]
    \centering
    \includegraphics[width=\linewidth]{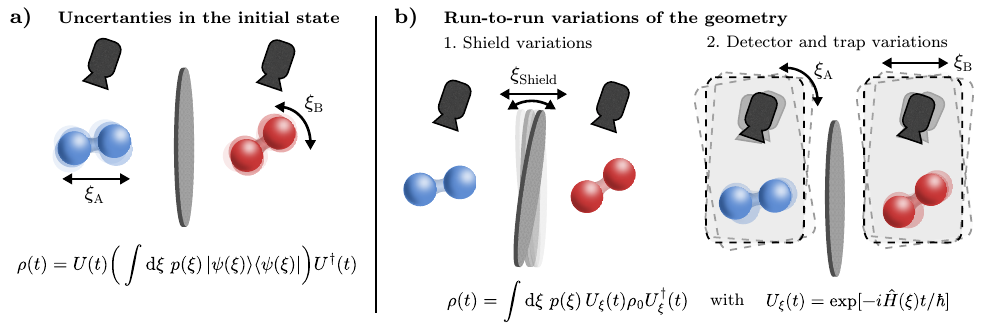}
    \caption{\justifying\small
    \textbf{Geometry variations}
    \textbf{a,} Uncertainty in the preparation of the initial state. The initial state $\rho(0) = \int \dd \xi \, p(x) \ketbra{\psi(t)}{\psi(t)}$ is mixed. 
    The time evolution does not generate additional decoherence and any loss of purity is solely inherited from the initial state. 
    \textbf{b,} Run-to-run variations of the setup. The varying geometry results in run-dependent time evolution operators $U_\xi(t)$ which generate, after averaging over $\xi$, decoherence even if the initial state $\rho_0$ is pure.
    We distinguish the two experimentally relevant cases of distance $\xi^{(L)}$ and orientation $\xi^{(\theta)}$ variations in the orientation and placement of the \textit{1. shield} and \textit{2. detectors/traps} discussed in the text.}
    \label{fig:geometry-variations}
\end{figure*}

In the lab frame, all relevant objects -- the two local potentials, the two measuring devices and the shield -- may undergo stochastic displacements and rotations relative to their desired position and orientation.
We model these fluctuations by independent Gaussian variables $\xi_i\sim\mathcal{N}(0,\Delta \xi^2)$. 
These fluctuations may or may not be correlated across different objects, depending on the specifics of the experimental configuration.
In the following, we describe the state of the system in the detector frame, which defines the operational frame in which the density matrix is reconstructed.
It is useful to distinguish two qualitatively different types of stochasticity:
\begin{enumerate}
    \item[\textbf{a)}] \textbf{Uncertainty in initial state} \\
    In this case, randomness enters exclusively through the initial state, while the time evolution is governed by the same Hamiltonian and therefore identical for all realizations.
    The dynamics is therefore described by a single unitary channel acting on a classical mixture
        \begin{equation*}
        \rho(t) =  U(t) \left( \int \dd \xi \, p(\xi) \ketbra{\psi(\xi)}{\psi(\xi)} \right) U^\dagger(t).
    \end{equation*}
    As a consequence, no additional decoherence is generated during the evolution; any loss of purity is solely inherited from the initial state.
    
    \item[\textbf{b)}] \textbf{Run-to-run variation of the setup} \\
    Here, fluctuations in the experimental arrangement lead to different dynamics in each run, resulting in an ensemble-averaged state at time $t$,
    \begin{equation}
        \rho(t) = \int \dd \xi \, p(\xi) \, U_{\xi}(t) \rho_0 U_{\xi}^\dagger(t),
    \end{equation}
    where $U_{\xi}(t) = \exp[-i \op H(\xi)t/\hbar]$.
    The dynamics is thus described by a convex mixture of unitary channels and gives rise to decoherence.

    We consider two experimentally relevant instances that lead to two different forms of noise, both visualized in \figref{fig:geometry-variations}(b):
    \begin{enumerate}
    \item[\textit{1.}] \textit{Shield variations:}
        The two detectors and two traps are all fixed to the lab frame, while the shield undergoes stochastic displacements $\xi^{(L)}_\mathrm{Shield}$ and rotations $\xi^{(\theta)}_\mathrm{Shield}$ that affect the Hamiltonian observed by the particles in each realization $H(\xi)$.
        
    \item[\textit{2.}] \textit{Detector variations:} 
        Each trapping potential is rigidly attached to its local detector, but, the distance and orientation of each detector relative to the shield and to each other fluctuates.
        In this case each detector observes a different fluctuation of the shield's position and orientation, $\xi^{(L)}_{A/B}\sim\mathcal{N}(0,\Delta L^2)$ and $\xi^{(\theta)}_{A/B}\sim\mathcal{N}(0,\Delta\theta^2)$, respectively, where the fluctuations are, in general, uncorrelated across the two detectors.
        Such a situation may be found, for example, where the same elements are used for trapping and for detection, or when the detector always measures displacements with respect to the equilibrium position of the trap.
    \end{enumerate}
\end{enumerate}
We note that, although both, initial-state uncertainty and run-to-run variations of the experimental arrangement, induce fluctuations in the particle--shield separation, they are not equivalent, and lead to qualitatively different forms of noise.
The case where the trapping potentials fluctuate relative to the detectors leads to decoherence also in the absence of particle--shield interactions and is discussed separately in Sec.~\ref{sec:trap-stability}. 

For small spatial delocalizations $\Delta x, \Delta y \ll L$ as well as small geometric variations $\Delta L \ll L$ and $\Delta \theta \ll L / \Delta x$, it is well justified to expand the Hamiltonian $\op H(\xi)$ up to second order in the position operators and in the stochastic parameters $\xi$.
The explicit form of these expansions is given in Appendices~\ref{apx:interactions-casimir} and~\ref{apx:interactions-mag-dipole}.

For the parameter regimes and electrically neutral particles considered here, coherence loss is dominated by Casimir and magnetic-dipole interactions of each particle with the shield. 
Fluctuations in the gravitational coupling are therefore negligible.
While an analytical justification is provided in Appendix~\ref{apx:interactions-gravity}, this approximation has also been verified numerically. 

In the following, we neglect the trapping potential in the Hamiltonian during the entanglement generation step, i.e. setting $\omega = 0$. 
During the preparation and detecting stage in the GIE protocols, a trap may however still be employed.
For evolution times $t \lesssim 2\pi/\omega$ the harmonic confinement does not significantly modify the dynamics, while its omission considerably simplifies the analysis. 
Moreover, the bounds on admissible setup variations derived below remain largely unaffected by the inclusion of a weak trapping potential; numerical results indicate that they become slightly more stringent.

\subsection{Gaussian states}\label{subsec:gauss-variations}
We consider squeezed Gaussian states delocalized along the $x$-direction and $y$-direction, modeled as harmonic oscillator modes with position widths $\Delta x$ and $\Delta y$.
Gaussian states are fully characterized by their first and second moments, given by the displacement vector $\mathbf{d} = \mean{\mathbf{r}}$ and the second-moment matrix $V_{ij} = \tfrac{1}{2}\mean{r_i r_j + r_j r_i}$, where $\mathbf{r} = (X_1, P_1, \dots, X_n, P_n)^\intercal$ collects the quadratures.
The covariance matrix is then defined as $\sigma = V - \mathbf{d}\mathbf{d}^\intercal$.
For convenience, we introduce dimensionless quadratures $X_i \equiv \op{x}_i/x_0$ and $P_i \equiv \op{p}_i/p_0$ for some $x_0$ and $p_0 = \hbar/x_0$ such that $[X_i,P_j] = i \delta_{ij}$.
These satisfy the uncertainty relation $\Delta X_i \Delta P_i \geq \tfrac{1}{2}$.
Initially, all modes $m$ are in a thermal state
\begin{align}\label{eq:gaussian-initial-state-d}
    \mathbf{d}(0) &= 0 \qquad\text{and} \\
     V(0) = \!\!\!\!\!\!\! \bigoplus_{\substack{m=\{X_A,X_B\}}} \!\!\!\!\!\!(1+&2\bar{n}_m)\diag\left(\Delta m^2, \Delta P_m^2\right), \label{eq:gaussian-initial-state-V}
\end{align}
where $\bar n_m$ denotes the mean thermal occupation of mode $m$.
The initial state is pure when $\bar{n}_m=0$ for all $m$.
We always take $\bar n_y = 0$ in the following.

Under a quadratic Hamiltonian $\op{H} = \tfrac{1}{2}\mathbf{r}^\intercal G \mathbf{r} + \mathbf{g}^\intercal \mathbf{r}$, Gaussianity is preserved at all times.
The  time evolution of $V(t)$ and $\mathbf{d}(t)$ follows from the Heisenberg equations of motion and are shown in Appendix~\ref{apx:time-evolution-gaussian-formalism}.
For each realization $\xi(t)$, the system evolves to a different Gaussian state, such that the ensemble-average covariance matrix is given by
\begin{equation}\label{eq:mean-covariance-matrix}
    \bar{\sigma}(t) = \mean{V(t)}_\xi - \mean{\mathbf{d}(t)}_\xi\mean{\mathbf{d}(t)}_\xi^\intercal.
\end{equation}

Equivalently, this can be expressed as
\begin{equation*}
    \bar{\sigma} (t) = \mean{\sigma(t)}_\xi + \Cov_\xi[\mathbf{d}(t)],
\end{equation*}
where $\Cov_\xi[\mathbf{d}] = \mean{\mathbf{dd}^\intercal}_\xi - \mean{\mathbf{d}}_\xi\mean{\mathbf{d}}_\xi^\intercal$ represents the covariance of the fluctuating mean displacements.
This expression highlights that averaging only over specific realizations of $\sigma_\xi(t)$ would underestimate the total variance of the averaged state: 
random shifts of the mean displacement increase the overall variance, even if each realization $\sigma_\xi(t)$ exhibits the same intrinsic quantum noise. 

To quantify entanglement between two subsystems with $m$ and $n$ modes, we use the logarithmic negativity~\cite{Plenio2005}, which for Gaussian states is given by 
\begin{equation}\label{eq:log-neg-gaussian}
E_N = \sum_{j=1}^{m+n} \max\{0, -\log_2 (2\nu_j)\} .
\end{equation}
Here, $\nu_j$ are the $m+n$ symplectic eigenvalues of the covariance matrix $\sigma^\Gamma$ of the partially transposed state, obtained by applying the phase space transformation $p_i \mapsto -p_i$ to $\sigma$ for all the modes $i$ in one subsystem~\cite{Serafini2017}.

We first analyze entanglement generation in the presence of only gravitational interaction and no variations of the setup, providing a baseline for comparison with the case where Casimir and magnetic-dipole interactions together with setup fluctuations are included.
By expanding the gravitational Hamiltonian to second order in the particle position operators~\cite{Pedernales2023}, we obtain the entanglement rate $\Gamma_\mathrm{ent.} = \partial_t E_N$ as $t\rightarrow 0$  in Appendix~\ref{apx:entanglement-gaussian} for $\theta_A=\theta_B\equiv \theta$ by ar
\begin{multline}\label{eq:gaussian-entanglement-rate}
    \Gamma_\mathrm{ent.} = \frac{\lambda \Delta X_A \Delta X_B x_0^2}{\hbar \log 2}\abs{2\sin^2\theta- \cos^2\theta} \\
    + \frac{\lambda \Delta Y_A \Delta Y_B x_0^2}{\hbar\log2}\abs{2\cos^2\theta - \sin^2\theta},
\end{multline}
with the gravitational coupling strength $\lambda=GM_AM_B / (4L^3)$.
For $\Delta X \gg \Delta Y$, this rate is maximized for $\theta = \pi/2$, when the delocalization axis is oriented perpendicular to the shield in the linear orientation and exceeds the parallel configuration by a factor of $2$.
This behavior is expected: in this configuration, the spatial delocalization maximizes the spread between the shortest and longest interparticle separations sampled by the wavefunction.
As a result, the gravitational interaction generates the largest differential phase accumulation across the support of the state, leading to a faster buildup of non-local correlations.

\begin{figure*}[!th]
    \centering
    \includegraphics[width=\linewidth]{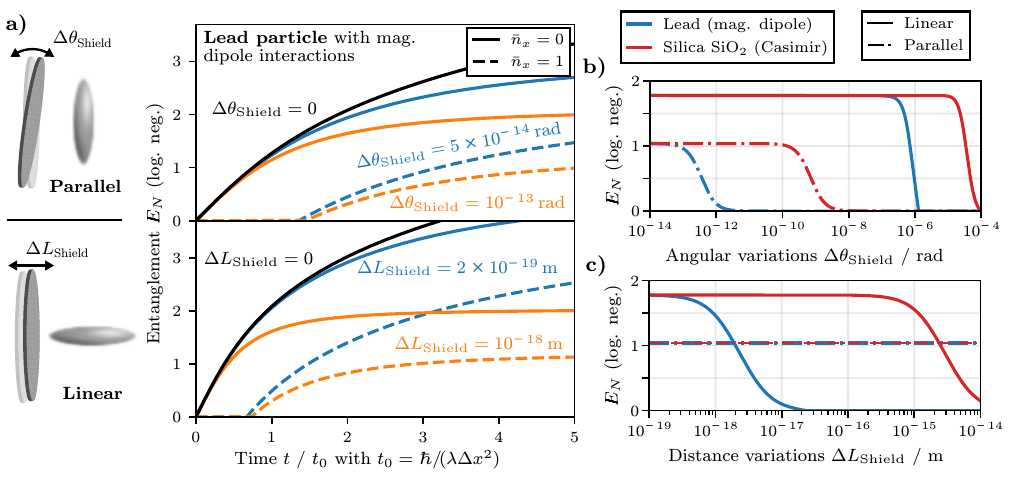}
    \caption{\justifying\small
    \textbf{Entanglement between Gaussian squeezed states subject to \textit{shield variations (see \figref{fig:geometry-variations}(b)1.)}}
    \textbf{a,} Time evolution of the logarithmic negativity $E_N$ for Gaussian wave packets in the presence of particle--shield magnetic dipolar interactions in the parallel ($\theta=0$, \textbf{top}) and linear ($\theta=\pi/2$, \textbf{bottom}) orientation.
    The parallel configuration is illustrated with run-to-run angular fluctuations of the shield orientation, while the linear orientation is shown with variations in the lateral shield placement.
    The principal $x$-mode is initially in a thermal state with occupation number $\bar n_x$.
    Time is expressed in units of $t_0 = \hbar/(\lambda \Delta x)$ with the gravitational coupling strength $\lambda = GM_AM_B / (4L^3)$, corresponding to $t_0\approx 164\si{ms}$ for silica particles and $t_0\approx 9\si{ms}$ for superconducting lead particles with the parameters of \tabref{tab:parameters}.
    For times $t/t_0 \lesssim \Delta x\Delta p / \hbar$, the entanglement grows linear as $E_N \propto t/t_0$~\cite{Pedernales2022}.
    \textbf{b,} Logarithmic negativity at $t=t_0\pi/4$ as a function of angular variations $\Delta \theta_\mathrm{Shield}$. 
    For lead particles in the linear configuration, the small but non-zero transverse $y$-mode has a visible effect on the stability, explaining the distinct look compared to other curves.
    \textbf{c,} Logarithmic negativity at the same time as a function of distance variations $\Delta L_\mathrm{Shield}$.}
    \label{fig:gauss-shield-variations}
\end{figure*}
In the following, we study the effect on entanglement generation of random variations in the setup geometry.
Note that, after averaging over the classical fluctuations, the resulting state is in general no longer guaranteed to be Gaussian.
Higher order cumulants become non-vanishing and the averaged state has to be written perturbatively as $\rho=\rho_G + \Delta \rho$, where $\rho_G$ denotes the Gaussian state described by the covariance matrix $\bar \sigma$.
The trace-norm $\norm{\Delta\rho}_1$ is bounded in Appendix~\ref{apx:non-gaussianity} and scales in the variations $\Delta \xi$ as $\norm{\Delta\rho}_1 \sim \Delta \xi^2 \norm{\rho_G}_1 + \mathcal{O}(\Delta \xi^4)$.
The deviation of the logarithmic negativity satisfies $\Delta E_N = \abs{E_N(\rho_G)-E_N(\rho)} \leq \norm{\Delta \rho^\Gamma}_1/\norm{\rho_G^\Gamma}_1$, where $\rho^\Gamma$ is the partial transposition of a state $\rho$ with respect to any subsystem.
For small variations $\Delta \xi$, the state is therefore almost Gaussian with $\lim_{\Delta \xi \rightarrow 0} \Delta E_N = 0$, and the logarithmic negativity of $\rho_G$ serves as a good estimator of the logarithmic negativity of $\rho$.

\subsubsection{Variations in the shield positioning}
If position and orientation of the shield fluctuate slightly from run to run, the particle--shield interaction Hamiltonian changes accordingly as
\begin{equation*}
    \op H(\xi) = \op H_\mathrm{Grav.} + \op H_\mathrm{Cas./mag.\,Dip.}(\xi_\mathrm{Shield}).
\end{equation*}
Expanding the Hamiltonian up to second order in both, the stochastic variables $\xi_\mathrm{Shield}$ and the position operators, allows us to analytically evolve and average the resulting state.
The full expression of the expanded interaction Hamiltonians for both considered orientations are given in Appendices~\ref{apx:interactions-gravity} to~\ref{apx:interactions-mag-dipole}.
The logarithmic negativity as a function of the standard deviation of the fluctuations $\Delta L_\mathrm{Shield}$ and $\Delta \theta_\mathrm{Shield}$ is shown in \figref{fig:gauss-shield-variations} for both pure and thermal initial states.

An initial state with thermal population $\bar n_x > 0$ results in a delayed onset of entanglement growth, in alignment with Ref.~\cite{krisnanda2020}.
The entanglement generation and its sensitivity to variations depends strongly on the orientation of the states with respect to the shield.

In the parallel orientation, entanglement grows more slowly but is more robust against fluctuations in the particle--shield separation $L$, since all components of the delocalized wave packet acquire the same random phase.
Angular variations however, rapidly suppress entanglement once $\Delta \theta_\mathrm{Shield}$ exceeds approximately $10^{-9}\si{rad}$ for Casimir interactions between silica particles and $5\times10^{-13}\si{rad}$ for superconducting lead dominated by magnetic-dipole interactions.

In contrast, for the linear configuration the entanglement initially grows twice as fast and is significantly more tolerant to angular misalignment.
Variations in the separation, however, suppress entanglement for fluctuations exceeding $\Delta L_\mathrm{Casimir} \approx 10^{-14}\si{m}$ and $\Delta L_\mathrm{mag.\,Dipole} \approx 10^{-17}\si{m}$.
\begin{figure*}[!th]
    \centering
    \includegraphics[width=\linewidth]{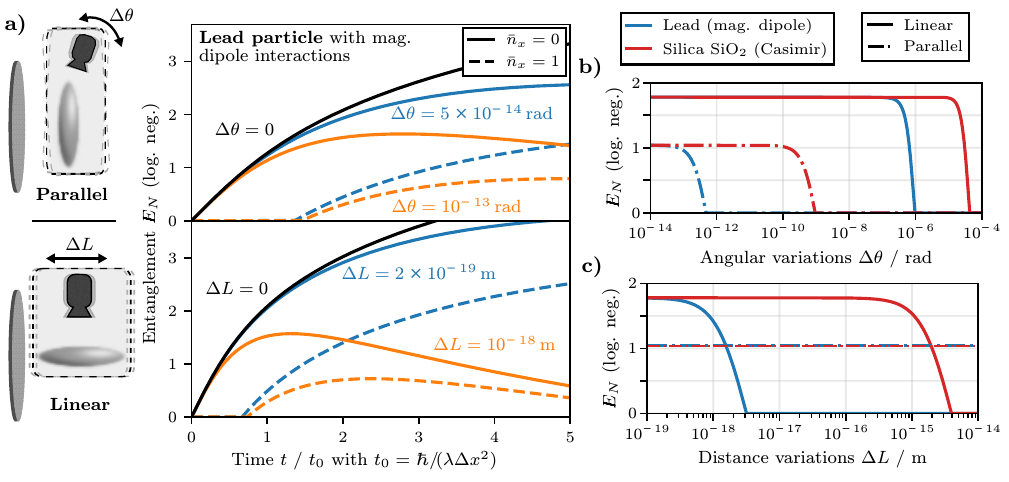}
    \caption{\justifying\small
    \textbf{Entanglement between Gaussian squeezed states subject to \textit{detector and trap variations (see \figref{fig:geometry-variations}(b)2.)}}
    \textbf{a,} Time evolution of the logarithmic negativity $E_N$ for Gaussian wave packets in presence of particle--shield magnetic dipolar interactions in the parallel (\textbf{top}) and linear (\textbf{bottom}) orientations.
    The parallel configuration is illustrated with angular fluctuations of the detectors over many runs, while the linear configuration is shown with variations in the detector positioning.
    After saturation, the entanglement decays much faster back to zero compared to the previous case.
    \textbf{b,} Logarithmic negativity at time $t=t_0\pi/4$ as a function of angular variations $\Delta \theta$ for both orientations and both particle types.
    The difference in the shape of the curves compared to \figref{fig:gauss-shield-variations} stems from the uncorrelated nature of the variations.
    \textbf{c,} Logarithmic negativity at the same time as a function of placement fluctuations $\Delta L$ relative to the static shield.
    }
    \label{fig:gauss-detector-variations}
\end{figure*}
The linear configuration has an additional practical limitation that persists even under ideal setup alignment.
Because different parts of the spatial superposition are located at different distances from the shield, they experience substantially different Casimir or magnetic-dipole potentials.
The resulting energy difference $\Delta V=V(\Delta x)-V(-\Delta x)$ leads to a rapid relative phase accumulation at a rate $\Gamma \sim |\Delta V|/\hbar$.
For the parameters in \tabref{tab:parameters}, this corresponds to $10.4\si{GHz}$ for a silica particle experiencing Casimir interactions and $378\si{THz}$ for a superconducting lead particle experiencing magnetic-dipole interactions.
As a consequence, the entanglement readout becomes extremely sensitive to the precise measurement time:
The local phases evolve on sub-nanosecond timescales, so even small timing uncertainties translate into large phase uncertainties. 
Although this unitary dynamics preserves entanglement, averaging over states sampled at slightly different times, hence with different local phases, is equivalent to local dephasing and suppresses the coherences responsible for entanglement.
Resolving the gravitationally induced correlations therefore requires sub-nanosecond timing control to sample the state at a well-defined relative phase.

\subsubsection{Variations in the position of the detectors and traps}
Variations in the positioning of the detectors and traps relative to each other and to the shield can be modeled by including the uncorrelated local fluctuations $\xi^{(L)}_{A/B}$ and $\xi^{(\theta)}_{A/B}$ directly in the interaction Hamiltonian as
\begin{equation*}
    \op H(\xi) = \op H_\mathrm{Grav.} + \op H_\mathrm{Cas./mag.\,Dip.}(\xi_A, \xi_B) .
\end{equation*}
The previous case of shield variations can actually be expressed by using the same fluctuations in a correlated manner, such that $\xi_A^{(L)} = - \xi_B^{(L)} \equiv \xi^{(L)}_\mathrm{Shield}$ and $\xi_A^{(\theta)} = \xi_B^{(\theta)} \equiv \xi^{(\theta)}_\mathrm{Shield}$.

The resulting entanglement dynamics are shown in \figref{fig:gauss-detector-variations}.
Compared to the previous case discussed above, the system is now more sensitive to fluctuations, with tolerance thresholds reducing by nearly one order of magnitude.

\subsection{Two-level cat-states}\label{subsec:cat-variations}
The second class of initial states we investigate here are massive Schrödinger cat-states, consisting of symmetric superpositions of spatially separated states.
The combined initial state of the two particles is then written as 
\begin{equation*}
    \ket{\psi_\mathrm{Cat}} = \frac{1}{2}
    \left(\big|\psi^{(1)}\big\rangle + \big|\psi^{(2)}\big\rangle\right)_A
    \otimes
    \left(\big|\psi^{(1)}\big\rangle+\big|\psi^{(2)}\big\rangle\right)_B.
\end{equation*}
We constrain ourselves to the regime in which the states $\left\{\ket{\psi^{(1)}},\ket{\psi^{(2)}}\right\}$ can be approximated as orthonormal and therefore span an effective two-dimensional Hilbert space.
This corresponds to the limit in which the spatial overlap $\braket{\psi^{(1)}}{\psi^{(2)}} = 0$ vanishes due to a sufficiently large spatial separation of the wave packets.
In this regime, the motional degree of freedom of each particle can be approximated by a two-level system encoding the two branches of the spatial superposition.
We refer to these states as \q{two-level cat-states}.

Within this effective description, the continuous position operators can be discretized:
If the delocalization satisfies $\Delta x \gg \Delta y$, the operators reduce to $\op{x}\rightarrow\Delta x \sigma_z$ and $\op{y}\rightarrow0$, where $\sigma_z$ is the Pauli matrix.
For general orientations $\theta_{A/B}$, the time evolution of the entanglement under the gravitational interaction is derived in Appendix~\ref{apx:cat-states-gravity} and is given by
\begin{equation}\label{eq:cat-state-entanglement}
    E_N=\log_2\left(1+\abs{\sin(\phi t)}\right),
\end{equation}
with
\begin{equation*}
    \phi = \frac{\lambda \Delta x_A \Delta x_B}{\hbar} \left(2\sin\theta_A\sin\theta_B-\cos\theta_A\cos\theta_B\right).
\end{equation*}
As in the Gaussian case, entanglement growth is maximal in the linear configuration.

\subsubsection{Variations in the shield positioning}
Here, we again consider the stability of entanglement against fluctuations of the shield position and orientation, while all other geometries stay fixed.

The resulting stability of the entanglement between cat-states with respect to variations $\Delta L_\mathrm{Shield}$ and $\Delta \theta_\mathrm{Shield}$ is shown in \figref{fig:cat-shield-variations}. 
In addition, the first-order analytical prediction derived in Appendix~\ref{apx:cat-states-variations} is shown in gray.
\begin{figure}[!t]
    \centering
    \includegraphics[width=\linewidth]{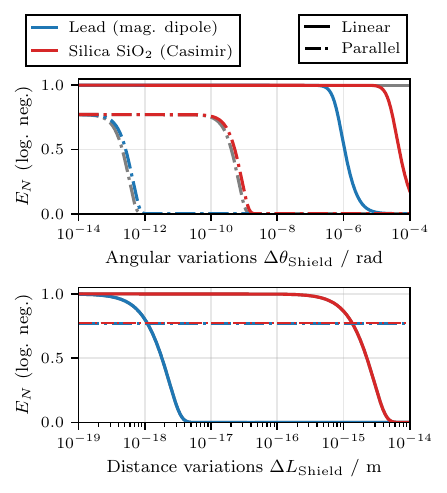}
    \caption{\justifying\small
    \textbf{Entanglement for two-level cat-states subject to \textit{shield variations (see \figref{fig:geometry-variations}(b)1.)}}
    Logarithmic negativity in the parallel and linear orientation after evolution times $t=t_0\pi/4$, corresponding to $t\approx 129\si{ms}$ for silica particles and $t\approx 7 \si{ms}$ for superconducting lead.
    We use the parameters from \tabref{tab:parameters}.
    \textbf{top,} Entanglement as a function of fluctuations in the shield orientation. 
    In gray we show the first-order analytical expressions derived in Apx.~\ref{apx:cat-states-variations}.
    \textbf{bottom,} Entanglement as a function of fluctuations in the particle--shield separation.
    The qualitative dependence on orientation and type of the variation closely mirrors the behavior for Gaussian states in \figref{fig:gauss-shield-variations}.}
    \label{fig:cat-shield-variations}
\end{figure}
As before, the robustness of entanglement depends strongly on the particle orientation: 
In the parallel configuration ($\theta = 0$), decoherence arises predominantly from angular fluctuations, whereas in the linear configuration ($\theta = \pi/2$), entanglement is sensitive to variations in the particle--shield separation.

The qualitative behavior closely matches that observed for Gaussian states.
The apparently enhanced robustness of cat states against angular fluctuations in the linear configuration is not intrinsic, but results from the inclusion of a finite transverse $y$-mode in the Gaussian analysis.
In the limit $\Delta y\rightarrow 0$, both state classes exhibit nearly identical stability thresholds, with Gaussian states showing a marginally higher robustness.

\subsubsection{Variations in the position of the detectors and traps}
The effect of variations in the placement of the detectors can be treated analytically in a compact manner, as detailed in Appendix~\ref{apx:cat-states-variations}.
To first order in $\xi^{(j)}$ with $j\in\{L, \theta\}$, the logarithmic negativity between two identically prepared particles ($\theta_A = \theta_B \equiv \theta$ and $\Delta x_A = \Delta x_B \equiv \Delta x$), is given by
\begin{equation}\label{eq:log-neg-cat-variations}
    E_N = \max\left\{0,\,\log_2\left[e^{-\gamma^2} \left(\cosh(\gamma^2) + \abs{\sin(\phi t)}\right)\right]\right\} ,
\end{equation}
with $\gamma^2 = \sum_{j\in\{L,\theta\}} (\gamma^{(j)}_\mathrm{Casimir})^2 + (\gamma^{(j)}_\mathrm{mag.\,Dipole})^2$.
For Casimir-dominated particle--shield interactions, $\gamma_\mathrm{Casimir}$ take the form
\begin{align*}
    \gamma^{(\theta)}_\mathrm{Casimir} &= \sqrt{8}\, (L-R-d_s/2)\,\eta\,t\, \Delta x\, \Delta\theta\, \cos\theta\,/\,\hbar, \\
    \gamma^{(L)}_\mathrm{Casimir} &= \sqrt{72}\, \eta\, t\, \Delta x\, \Delta L\, \sin\theta\,/\,\hbar,
\end{align*}
where $\eta=\hbar c \pi^3 R \vphi(\veps_r) / (720(L-R-d_s/2)^4)$ denotes the Casimir coupling strength.
For superconducting lead particles, where particle--shield interactions are dominated by induced magnetic-dipole moments, we find
\begin{align*}
    \gamma^{(\theta)}_\mathrm{mag.\,dipole} &= \sqrt{18}\, (L-d_s/2)\,\delta\,t\, \Delta x\, \Delta\theta\, \cos\theta\,/\,\hbar, \\
    \gamma^{(L)}_\mathrm{mag.\,dipole} &= \sqrt{288}\,\delta\, t\, \Delta x\, \Delta L\, \sin\theta\,/\,\hbar,
\end{align*}
with $\delta = 2\mu_0\abs{\vec{m}}^2 / (32\pi (L-d_s/2)^5)$.

\begin{figure*}[!th]
    \centering
    \includegraphics[width=\linewidth]{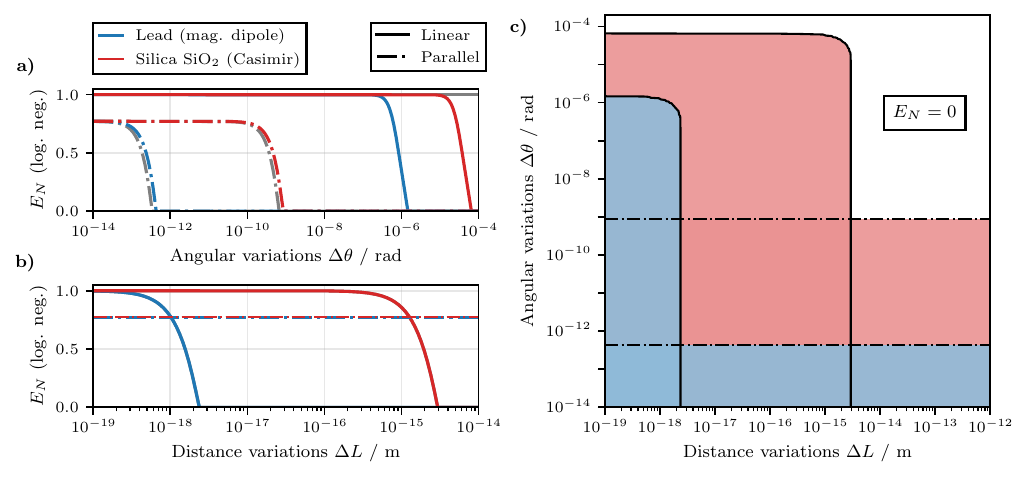}
    \caption{\justifying\small
    \textbf{Entanglement for two-level cat-states subject to \textit{detector and trap variations (\figref{fig:geometry-variations}(b)2.)}}
    Logarithmic negativity in the parallel and linear orientation after evolution times $t=t_0\pi/4$.
    We use the parameters from \tabref{tab:parameters}.
    \textbf{a,} Entanglement as a function of angular variations with the first-order analytical expressions of \Eqref{eq:log-neg-cat-variations} shown in gray.
    \textbf{b,} Entanglement as a function of fluctuations in the particle--shield separation in analogy to \figref{fig:gauss-detector-variations}(c).
    \textbf{c,} Entanglement for combined variations $\xi^{(L)}_{A/B}$ and $\xi^{(\theta)}_{A/B}$ in both discussed orientations (black outlines) after an evolution time of $t=t_0 \pi/4$.
    In the white region, the logarithmic negativity is zero.}
    \label{fig:cat-detector-variations}
\end{figure*}

In \figref{fig:cat-detector-variations}, we present the full numerical results and the analytical prediction from \Eqref{eq:log-neg-cat-variations}.
Compared to the shield-variations shown in \figref{fig:cat-shield-variations}, the system exhibits a reduced robustness against these uncorrelated fluctuations, mimicking the results obtained for Gaussian states discussed in Sec.~\ref{subsec:gauss-variations}.

\subsection{Continuous cat-states}\label{subsec:continuous-cat-variations}
\begin{figure*}[!thb]
    \centering
    \includegraphics[width=\linewidth]{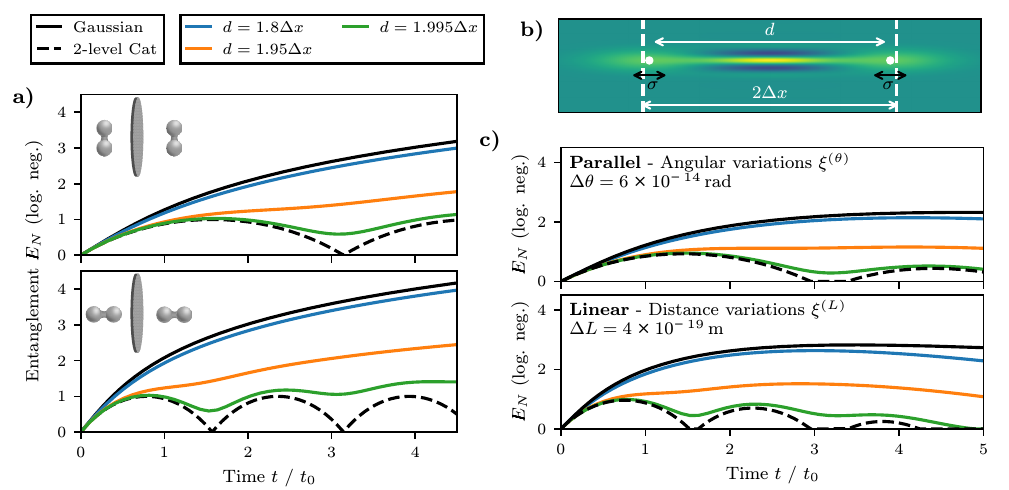}
    \caption{\justifying\small\textbf{Entanglement dynamics for continuous cat-states.}
    \textbf{a,}
    Logarithmic negativity $E_N$ for continuous initial states interpolating between squeezed Gaussian states and two-level cat-states, computed numerically in a truncated Fock basis at zero temperature.
    Results are shown for the parallel (\textbf{top}) and linear (\textbf{bottom}) orientations with variations in detector positioning.
    For reference, the corresponding analytical solutions for Gaussian and two-level cat-states with identical position variance $(\Delta x)^2$ are shown in black.
    The initial linear growth of entanglement with rate $\Gamma_\mathrm{ent.}=\partial_t E_N |_{t=0}$ is identical across the family of states; in agreement with Ref.~\cite{Pedernales2023}.
    \textbf{b,} Continuous cat-state in analogy to \Eqref{eq:continuous-cat} with positional standard deviation $\Delta x$, where the parameter $d\in[0,2\Delta x]$ interpolates between a fully Gaussian state ($d=0$ and $\sigma=\Delta x$) and a two-level cat-state ($d\rightarrow 2\Delta x$ and $\sigma\rightarrow0$).
    \textbf{c,} Entanglement dynamics including magnetic-dipole coupling of superconducting lead particles to the shield.
    We show representative angular variations (parallel orientation, \textbf{top}) and separation variations (linear orientation, \textbf{bottom}), evaluated numerically using Gauss-Hermite quadratures.
    As in panel \textbf{a}, the behavior interpolates continuously between the Gaussian and two-level cat-state limits.}
    \label{fig:entanglement-dynamics-continuous-cats}
\end{figure*}
Gaussian states and two-level cat-states form limiting cases of a continuous family of delocalized states. 
Establishing this connection enables a unified comparison of their entanglement generation.
We consider the symmetric superposition
\begin{equation}\label{eq:continuous-cat}
    \ket{\psi} = \frac{1}{\mathcal{N}} \left(\ket{\psi_+} + \ket{\psi_-}\right)_A \otimes \left(\ket{\psi_+} + \ket{\psi_-}\right)_B
\end{equation}
where each branch is a displaced squeezed state $\ket{\psi_\pm} = D(\pm\alpha)S(r)\ket{0}$ with the position-space representation
\begin{equation*}
    \braket{x}{\psi_\pm}=\frac{1}{(2\pi \sigma^2)^{1/4}} \exp{-\frac{(x\pm d/2)^2}{4\sigma^2}} .
\end{equation*}
Here, $\sigma$ quantifies the width of each wave packet and $d$ their spatial separation.
The squeezing and displacement parameters are related by $r = -\log(\sigma / x_0)$ and a real displacement $\alpha = d / (4x_0)$ for an arbitrary scaling $\op{x} = x_0(\op{a} + \op{a}^\dagger)$.
\\
To quantitatively compare the entanglement generation of these generalized cat-states with the Gaussian and two-level limits discussed previously, we fix their position variance as
\begin{equation*}
    (\Delta x)^2 = \sigma^2 + \frac{d^2}{4(1+S)} ,
\end{equation*}
where $S = \braket{\psi_+}{\psi_-} = \exp\{-d^2/(8\sigma^2)\}$ is the overlap between the two branches. 

For $d = 0$, the state reduces to a single Gaussian wave packet with $\sigma = \Delta x$, recovering the Gaussian limit of Sec.~\ref{subsec:gauss-variations}.  
Conversely, for $d \to 2\Delta x$ (implying $\sigma \to 0$ and $S \to 0$), the state approaches the two-level cat-states analyzed previously.

The time evolution of this state is obtained numerically in a truncated Fock basis.
Ensemble averaging over the experimental parameters $\xi^{(L)}$ and $\xi^{(\theta)}$ is performed using exponentially convergent Gauss-Hermite quadratures.

As shown in \figref{fig:entanglement-dynamics-continuous-cats} for the case of variations in detector-positioning, the entanglement dynamics interpolate smoothly between the Gaussian and two-level cat-state regimes.
This confirms that the two classes of states are continuously connected.
\section{Trap stability}\label{sec:trap-stability}
We model fluctuations of the trapping potential by a stochastic force acting on each particle.
The corresponding Hamiltonian is given by
\begin{equation}\label{eq:hamiltonian-trap}
    \op H_\mathrm{Trap} = \sum_{i=A,B} \left( \frac{1}{2} M \omega^2 \op x_i^2 + \xi_i(t) \op x_i \right) ,
\end{equation}
where $\omega$ denotes the trap frequency and $\xi_i(t)$ are real-valued stochastic processes with units of force. They capture, for instance, slow drifts of the trap center, fluctuating forces due to residual gas collisions, or attraction to the shield.
Their effect is to shift the equilibrium position of the trapping potential by a distance $\xi^{(L)}_i(t) = \xi_i(t)/(M\omega^2)$.
We limit ourselves to zero-mean, stationary noise that is uncorrelated across particles,
\begin{equation*}
    \mean{\xi_i(t)} = 0, \qquad \mean{\xi_i(t)\xi_j(s)} = \delta_{ij} C_i(t-s),
\end{equation*}
with exponentially decaying time correlations
\begin{equation*}
    C_i(\tau) = \Delta \xi^2 \, e^{-|\tau|/t_c},
\end{equation*}
where $\Delta \xi^2$ sets the noise strength and $t_c$ the correlation time.
Including gravitational interactions and the particle--shield couplings introduced in Sec.~\ref{sec:setup}, the full Hamiltonian reads
\begin{equation}\label{eq:trap-complete-hamiltonian}
    \op H = \sum_{i=A,B} \frac{\op p_i^2}{2M} + \op H_\mathrm{Trap} + \op H_\mathrm{Grav.} + \op H_\mathrm{Cas./mag.\,Dip.} .
\end{equation}

\subsection{Quasi-static limit}
In the limit of long correlation times $t_c \gg t$, the noise correlator can be approximated as constant over the duration of a single realization of the experiment,
$C_i(\tau)\simeq\Delta\xi^2$.
In this regime, the stochastic force is effectively time-independent within a single experimental realization, corresponding to a static displacement of the trap center, described by $H_\mathrm{Trap}(\op x_i + \xi_i^{(L)})$ with $\xi^{(L)}_i \sim \mathcal{N}(0,\Delta L^2)$.
The state of the system at a given time $t$ is thus described by an ensemble of states, each evolved with randomly shifted equilibrium positions.

\begin{figure*}[!ht]
    \centering
    \includegraphics[width=\linewidth]{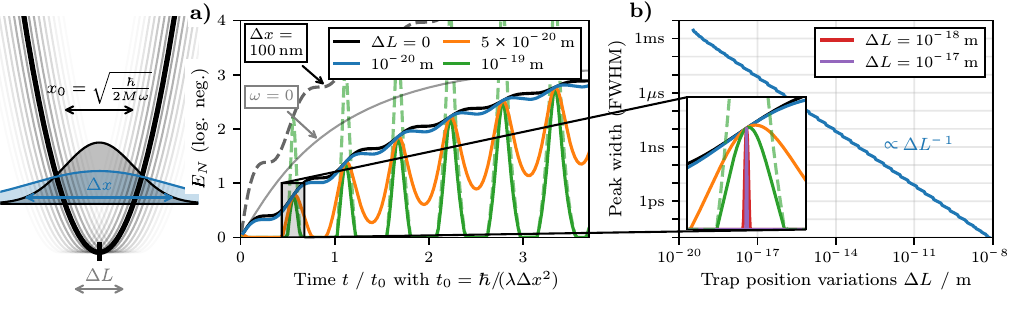}
    \caption{\small\justifying
    \textbf{Entanglement in the presence of quasi-static trap fluctuations.}
    \textbf{a,} Time evolution of the logarithmic negativity $E_N$ for magnetically levitated superconducting lead particles in an harmonic trap ($\omega = 100\times2\pi\si{Hz}$) in the linear configuration.
    Time is expressed in units of $t_0 = \hbar/(\lambda \Delta x^2)$ with the gravitational coupling strength $\lambda$, corresponding to $t_0 \approx 9\si{ms}$ for the superconducting lead particles shown here.
    Periodic oscillations arise from the harmonic motion of the trapped particles.
    Different run-to-run fluctuations $\Delta L$ of the trap center are included.
    Nonzero entanglement is observable only near integer multiples of $T/2=\pi/\omega'$.
    Continuous lines corresponds to $\Delta x = 50\si{nm}$, while dashed lines correspond to $\Delta x = 100\si{nm}$.
    \textbf{b,} Numerically determined peak-widths (FWHM) for different variations in seconds.
    We find, that $\text{FWHM}\propto 1/\Delta L$.}
    \label{fig:variations-trapping-potential}
\end{figure*}

Expanding the Hamiltonian \Eqref{eq:trap-complete-hamiltonian} to second order, the time evolution can be solved analytically within the Gaussian formalism (see Appendix~\ref{apx:gauss-trap-variations}).
The corresponding entanglement dynamics for the parameters listed in Table~\ref{tab:parameters} are shown in Fig.~\ref{fig:variations-trapping-potential}(a).
At short times, the entanglement grows linearly, while at later times, the dynamics exhibit periodic modulations due to the harmonic evolution of the trapped particles:
For each realization, the Gaussian state rotates around the point $(x_\mathrm{eq}, p_\mathrm{eq}) = (\xi^{(L)}, 0)$ in phase space, with an effective frequency
\begin{equation*}
    \omega' = \sqrt{\omega^2 - \frac{\lambda + 12\delta+6\eta}{M}},
\end{equation*}
where $\lambda$ is the gravitational coupling strength and $\delta,\eta$ account for magnetic and Casimir particle--shield interactions, respectively.

In the absence of noise, all realizations follow identical unitary dynamics, and the entanglement grows monotonically in time, although the entanglement rate $\Gamma_\mathrm{ent.} = \partial_t E_N$ exhibits periodic suppression while never vanishing exactly.
This is reflected in the recurrent near-flat segments of the black curve in \figref{fig:variations-trapping-potential}(a).
This behavior follows from $\Gamma_\mathrm{ent.} \propto \Delta x^2$~\cite{Pedernales2023} and from the time-dependence of the position variance
\begin{equation*}
    \Delta x(t) = \Delta x(0)\cos(\omega' t) + \frac{\Delta p(0)}{M\omega'}\sin(\omega' t).
\end{equation*}
For the strongly squeezed states, $\Delta x(0) \gg \Delta p(0)/(M\omega')$, the spatial width is minimized after a quarter period, $t = T/4$ with $T = 2\pi/\omega'$.
At this time, the wave packet spread in position reaches $\Delta x(T/4) = \hbar/(2M\omega'\Delta x(0))$.
As a consequence, the entanglement rate is large at the beginning of each cycle, strongly suppressed near odd multiples of $t=T/4$ and recovers again at later times.
In contrast, weakly squeezed states exhibit significantly smaller oscillations in the entanglement rate.

In the presence of fluctuations, $\xi^{(L)} \neq 0$, the entanglement grows no longer monotonically and becomes strongly suppressed, whenever the position variance is small in the course of the time evolution.
Averaging over random displacements leads to enhanced distinguishability of narrow wave packets, resulting in stronger decoherence and a corresponding reduction of entanglement.
Conversely, at times $n T/2$ ($n\in\mathbb{N}$), when the position variance is maximal, the effect of displacements is minimized and the entanglement approaches the noiseless value.

In such realistic scenarios, entanglement becomes detectable only within narrow time windows centered around multiples of $T/2$. 
We find that the size of these windows scales inversely with the amplitude of the fluctuations in the trap equilibrium position, i.e., $\propto 1/\Delta L$, as shown in \figref{fig:variations-trapping-potential}(b).

While stronger squeezing increases the achievable entanglement, it also enhances sensitivity to positional fluctuations.
Interestingly, the time window over which the entanglement remains non-zero is unaffected by the amount of squeezing and depends only on the fluctuation amplitude $\Delta L$.
Within this window, however, the maximum entanglement attained increases for more strongly squeezed states.
Consider, for example, a moderate trap frequency of $\omega = 100\times2\pi\si{Hz}$ and the parameters from \tabref{tab:parameters} for magnetically levitated lead particles, the position variance after a quarter of a period corresponds to an extremely small width $\Delta x(T/4) \sim 10^{-20}\si{m}$, which leads to an extreme sensitivity to static displacements at this point in time, with even minute variations of order $\Delta L \sim 10^{-20} \si{m}$ affecting the entanglement.

Reducing the squeezing mitigates this sensitivity but increases the entanglement generation time.
The minimum required squeezing is therefore set by the available coherence time $\tau$, the detectable entanglement threshold $E_{N,\,\mathrm{min}} \leq E_N(\tau)$, and the gravitational coupling $\lambda$.
Expressed relative to the trap's ground-state width $x_0=\sqrt{\hbar/(2M\omega)}$, the required delocalization of the two identical particles is given by
\begin{equation}
    \frac{\Delta x}{x_0} \geq \Delta x_\mathrm{min} \sqrt{\frac{2M\omega}{\hbar}} = \sqrt{\frac{3L_0^3\omega E_{N,\,\mathrm{min}}\log 2}{G\pi\rho\tau}}
\end{equation}
where $L_0 = L/R$ and $E_N(t)\approx 2t/(t_0 \log2)$, valid for small times $t \lesssim t_0$, was used in the last equality.
For representative parameters $\tau \sim 1\si{s}$, $E_{N\,\mathrm{min}}=10^{-2}$ and $L_0 = 1.5$ this corresponds to a squeezing parameter $r=-\log(\Delta x/x_0)\approx -8.3$, ($= 72\si{dB}$) which significantly exceeds currently achievable experimental squeezing levels~\cite{Kamba2025,Lei2016,Marocco2026}.

\subsection{Markovian limit}
The opposite regime of short correlation times $t_c \to 0$ corresponds to rapidly fluctuating forces.
To obtain a well-defined limit, the correlator must be rescaled such that its integrated strength remains finite.
We therefore write
\begin{equation*}
    C_i(\tau)=\frac{D}{t_c} \, e^{-\abs{\tau}/t_c} \xrightarrow{t_c\rightarrow 0}2D\delta(\tau) 
\end{equation*}
which effectively describes Gaussian white noise with diffusion constant $D$.

In this limit, the dynamics reduce to a Markovian equation for the covariance matrix (see Appendix~\ref{apx:gauss-trap-variations}),
\begin{equation*}
    \dv{\sigma}{t} = \Omega G \sigma + \sigma G^\intercal \Omega^\intercal + \mathcal{D},
\end{equation*}
where $\Omega$ denotes the symplectic form and $G$ is the quadratic Hamiltonian matrix associated with \Eqref{eq:trap-complete-hamiltonian}.
The diffusion matrix takes the form
\begin{equation*}
    \mathcal{D} = 2 D \bigoplus_{i=A,B}
    \begin{pmatrix}
        0 & 0 \\
        0 & 1
    \end{pmatrix},
\end{equation*}
corresponding to independent momentum diffusion acting on each mode.
As shown in Appendix~\ref{apx:gauss-trap-variations}, this evolution for the covariance matrix is equivalent to the Lindblad master equation
\begin{equation*}
    \dv{\rho}{t} = -\frac{i}{\hbar}[\op H,\rho] - \frac{D}{\hbar^2}\sum_{i=A,B}[\op x_i,[\op x_i,\rho]],
\end{equation*}
for the density matrix, which describes pure diffusion in position basis.
Physically, this represents the action of random kicks on the particle, changing its momentum and thus translates into incoherent broadening of its phase space distribution.

\begin{figure}[!th]
    \centering
    \includegraphics[width=\linewidth]{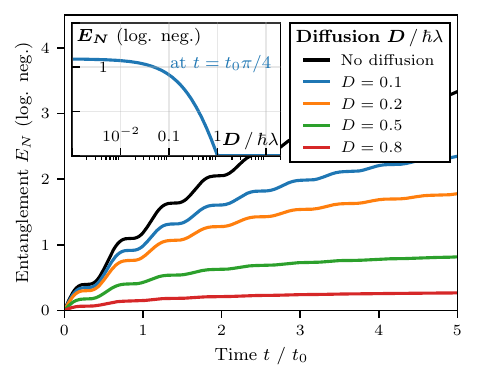}
    \caption{\small\justifying
    \textbf{Entanglement in the presence of Markovian trap fluctuations}
    Logarithmic negativity over time for different diffusion constants $D$.
    Here, $D$ is measured in units of $\hbar\lambda$ and for $D=1\hbar\lambda$, the entanglement rate $\Gamma_\mathrm{ent.}=\partial_tE_N$ is precisely equal to the decoherence rate.
    Consequently, for larger diffusion, no entanglement is observable, as seen in the inset plot.
    }
    \label{fig:markovian-limit}
\end{figure}

The entanglement dynamics are shown in \figref{fig:markovian-limit}.
In contrast to the quasi-static regime, the Markovian limit induces irreversible decoherence, and thus, no revivals of the entanglement are present.
While the Hamiltonian coupling generates inter-particle correlations, the local diffusion process degrades these correlations, leading to a suppression of the entanglement rate.
As a result, for sufficiently strong noise when the decoherence rate $\Gamma_\mathrm{dec.} = D\Delta x^2/\hbar^2$ is equal to the entanglement rate $\Gamma_\mathrm{ent.}\sim\lambda\Delta x^2/\hbar$, the dynamics become  entanglement-breaking, preventing the buildup of non-classical correlations despite the presence of coherent interactions.
\section{Quantum shield}\label{sec:shield}
Up to this point, the shield has been assumed to be a rigid plate with its position and orientation varying classically between runs.
Accordingly, this has been captured by fluctuating Hamiltonian parameters.
In the following, we now treat the shield as an explicit subsystem with its own quantum degrees of freedom coupled to those of the particles.
Its modes are assumed to be thermally populated and interact with the particles through Casimir or magnetic-dipole interactions, which, in general, contain both dispersive and dissipative components determined by the material response.
We focus on the regime of low temperature and low material losses, where dissipative contributions are strongly suppressed and the interaction can be approximated by an effective coherent Hamiltonian term.
In this regime, the shield may become entangled with the particles, thereby inducing decoherence, and additionally can also act as a coherent mediator of non-gravitational entanglement between them.
Such contributions are undesirable, as they may mask gravitationally induced entanglement if comparable in strength.
Since the goal of GIE experiments is to demonstrate entanglement mediated solely by gravity, any additional quantum channel would introduce a loophole in attributing the observed correlations to gravitational interaction.
It is therefore important to bound non-gravitational entanglement and identify regimes ensuring genuine GIE.
The mechanism of this entanglement mediation by the shield is similar in nature to previous findings of entanglement mediation by a massive object in a thermal state discussed in related contexts~\cite{Pedernales2022}.

We describe the shield as a circular elastic plate of thickness $d_s$ and radius $r_s$.
Its vibrational eigenmodes can be obtained analytically and are expressed in terms of Bessel functions.
Each mode is labeled by integers $(k,l)$ with $k \in [1,\infty)$, $l \in [0,\infty)$ and their spatial profiles are denoted by $u_{kl}(r,\vartheta,t)$ with corresponding resonance frequencies $\omega_{kl}$.
Explicit expressions are provided in Appendix~\ref{apx:interactions-shield} and representative mode shapes are shown in \figref{fig:shield-modes}.

Throughout the following analysis, we assume that the particle is positioned at the center of the shield $(r_0=0)$.
Small displacements against the center $r=r_0 \pm \xi^{(r)}$, however, do not affect the overall dynamics of the system considerably, as long as $\xi^{(r)} \ll r_s$.
For realistic shields with non-uniform mass distribution, the ideal particle placement will have to be found experimentally to minimize the particle--shield interactions.
We furthermore impose two conditions on the spatial extent of the particle wave packets:
(i) the delocalization along the principal axis, $\Delta x$, is small compared to the radial length scale on which $u_{kl}$ varies; and
(ii) the delocalization parallel to the shield, $\Delta z$, is sufficiently small so that the angular dependence of the mode can be neglected, i.e. $u_{kl}$ may be treated as effectively constant in $\vartheta$ over the support of the wave packet.

Both conditions are well satisfied in our parameter regime, as the lowest-order modes vary only on scales set by $r_s$ -- much larger than the extent of the particles' main delocalization.
Coupling to higher-order modes is strongly suppressed and therefore does not contribute appreciably to the dynamics as seen in the following section.

Under these conditions, the mode shape can be locally linearized around the particles' positions as $u_{kl}(r,\vartheta) \approx u_{kl}(0,\vartheta_0) + r \, \partial_r u_{kl}\big|_{r=0,\vartheta=\vartheta_0}$.
One can verify that the radial gradient $\partial_r u_{kl}\big|_{r=0}$ is maximal for $\vartheta_0=0$.
Hence, we adopt this configuration to obtain a worst-case estimate on the noise induced on the particles by the dynamical shield.

Each vibrational mode behaves as an independent quantum harmonic oscillator of frequency $\omega_{kl}$ and effective mass
\begin{equation*}
    m_\mathrm{eff} = \frac{m}{\pi r_s^2} 
    \int_\mathrm{Shield} \!\!r \dd r \dd \vartheta \, 
    \abs{u_{kl}(r,\vartheta)}^2,
\end{equation*}
where $m = \rho \pi r_s^2 d_s$ is the total mass of the shield.
Quantizing each mode leads to quadrature operators $\hat{q}_{kl}=\sqrt{\hbar/(2 m_\mathrm{eff} \omega_{kl})}(\hat{a}_{kl}^\dagger + \hat{a}_{kl})$ and $\hat{p}_{kl}=i \sqrt{\hbar m_\mathrm{eff} \omega_{kl}/2}(\hat{a}_{kl}^\dagger-\hat{a}_{kl})$, whose thermal variances at temperature $T$ fully describe the state and are given by
\begin{align}\label{eq:thermal-quadrature-variance}
    (\Delta q_{kl})^2 &= \frac{\hbar}{2m_\mathrm{eff}\omega_{kl}}\coth(\frac{\beta\hbar\omega_{kl}}{2}) 
    \quad \text{and} \quad \\
    (\Delta p_{kl})^2 &= \frac{m_\mathrm{eff} \hbar \omega_{kl}}{2}\coth(\frac{\beta\hbar\omega_{kl}}{2}) .
\end{align}
Here, $\beta=1/(k_B T)$ and $k_B$ is Boltzmann's constant. 
Equivalently, the thermal state of each mode can be written as
\begin{equation}\label{eq:thermal-state}
    \rho_{\mathrm{th},kl}
        = \int \mathrm{d}\alpha^2\, \frac{1}{\pi \bar{n}_{kl}}
          e^{-\frac{|\alpha|^2}{\bar{n}_{kl}}} \ketbra{\alpha} ,
\end{equation}
where the notation $\dd \alpha^2=\dd\Re\alpha\dd\Im\alpha$ is used and
$\bar{n}_{kl} = 1/(\mathrm{e}^{\beta \hbar \omega_{kl}} - 1)$ is the average thermal occupation number of mode $(k,l)$.
The full state of the shield is thus given by $\rho_\mathrm{Shield}=\bigotimes_{(k,l)}\rho_{\mathrm{th,}kl}$.

Two-level cat-states in front of the shield can therefore be initially described by the joint state $\rho_0 = \ketbra{\psi_\mathrm{Cat}}\otimes\rho_\mathrm{Shield}$.
The combined time evolution of particles together with the shield can be performed analytically and is provided in Appendix~\ref{apx:cat-states-shield}.

In the Gaussian-state formalism, we describe the shield modes by the quadrature set $(\{\op{q}_{kl},\op{p}_{kl}\})$. The initial covariance matrix of the two particles and the shield is then given by
\begin{equation*}
    \sigma = \sigma_\mathrm{Particles} 
    \oplus \bigoplus_{(k,l)} 
    \begin{pmatrix}
        (\Delta Q_\mathrm{kl})^2 & 0 \\
        0 & (\Delta P_\mathrm{kl})^2
    \end{pmatrix}
\end{equation*}
where $\sigma_\mathrm{Particles}$ is the covariance matrix for squeezed particles as in Sec.~\ref{subsec:gauss-variations}.

To extract the bipartite entanglement between both particles, we first must trace out the shield after the time evolution.
For the cat-states, this is done by performing the usual partial trace, whereas in the Gaussian formalism, one can simply discard all covariance-matrix elements associated with the shield modes~\cite{Serafini2017}.

\subsection{Decoherence effects of a quantum shield}
\begin{figure*}[!thb]
    \centering
    \includegraphics[width=\linewidth]{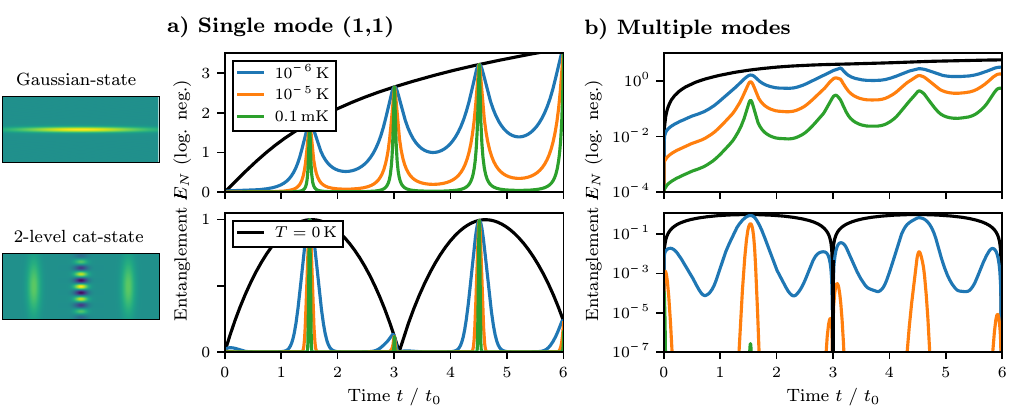}
    \caption{\justifying\small
    \textbf{Entanglement dynamics in a GIE setup in the presence of a quantum dynamical shield.}
    Results are shown for delocalized superconducting lead particles interacting with the shield via magnetic-dipole interactions prepared in the parallel orientation ($\theta = 0$), for which the gravitational signal is most robust against noise.
    The parameters from \tabref{tab:parameters} were used with a Niobium shield of size $d_s=2\si{\mu m}$ and $r_s=1\si{cm}$.
    If the particle--shield couplings are dominated by Casimir interactions instead, the corresponding dynamics are shown in \figref{fig:apx:shield-parallel-casimir}.
    \textbf{a,} Entanglement dynamics for Gaussian states (\textbf{top}) and two-level cat-states (\textbf{bottom}) with the shield dynamics restricted to the dominant $(1,1)$-mode only.
    Symmetry dictates that this mode provides the leading contribution to shield-induced decoherence in the parallel configuration.
    Partial revivals of gravitationally generated entanglement occur near times $t=2\pi j / \omega_{(1,1)}$ ($j\in\mathbb{N}$), where the single mode periodically re-phases.
    The peak widths $\sqrt{\omega_{(1,1)}/T} \sim\sqrt{d_s/(r_s^2 \, T)}$ decrease with increasing temperature.
    \textbf{b,} Corresponding dynamics where multiple shield modes are included (here: first 64 modes). 
    Because the modes have incommensurate frequencies, no time exists at which all modes simultaneously return to their initial quadratures and hence, the particles never fully decouple from the shield.
    }
    \label{fig:shield-parallel-magnetic-dipole}
\end{figure*}
We analyze the decoherence effects due to the presence of a quantum shield interacting with the particles in GIE experiments.
We constrain our analysis to the parallel orientation only, as particles in the linear orientation are prone to entanglement mediated via the shield, as discussed in detail in the next section.
Although the linear orientation could in principle be treated analogously, for our parameters the non-gravitationally mediated entanglement would obscure the gravitational signal, making the parallel configuration favorable for assessing gravitationally induced entanglement.
To investigate the interplay between  gravity and shield dynamics, we consider the Hamiltonian
\begin{equation}
    \op H=\op H_\mathrm{Gravity} + \op H_\mathrm{Shield}
\end{equation}
where the shield dynamics are governed by
\begin{equation}\label{eq:hamiltonian-shield}
    \op H_\mathrm{Shield} = \sum_{(k,l)}\left[\op H_{kl} + \hbar\omega_{kl}\left(\op a_{kl}^\dagger\op a_{kl} + \tfrac{1}{2}\right)\right]
\end{equation}
with $\op H_{kl}$ describing the Casimir- or magnetic-dipole coupling between the particles and the $(k,l)$ vibrational mode of the shield and is presented in Appendix~\ref{apx:interactions-shield}.
After linearizing the mode-profile $u_{kl}$ around the particles positions, $\op H_\mathrm{Shield}$ can be expanded to second order in its position operators $\{\op x_A, \op x_B, \op q_{kl}\}$, allowing us to solve the time evolution analytically in the case of two-level cat-states, as shown in Appendix~\ref{apx:cat-states-shield}.

Generally, the shield supports an infinite set of mechanical eigenmodes, however, only the lowest modes contribute considerably to shield-induced decoherences.
In the high-temperature limit $\hbar \omega_{kl}\ll k_B T$, which is well-satisfied for $T\gtrsim 10^{-9}\si{K}$ and typical mode frequencies $\omega\sim 50-200\si{s^{-1}}$, the variance of the oscillation amplitude scales as $\Delta q_{kl}\sim \omega_{kl}^{-1}$.
Therefore, low-frequency modes dominate the interaction.

The resulting entanglement dynamics for two-level cat-states and Gaussian states are shown in \figref{fig:shield-parallel-magnetic-dipole} and \figref{fig:apx:shield-parallel-casimir}.
One can see, that the quantum shield suppresses the bipartite entanglement generated by gravity.
Although a single mode would periodically decouple from the particles at times $t=2\pi j / \omega_{kl}$ ($j\in\mathbb{N}$), the shield hosts many modes with distinct and incommensurate frequencies. 
Consequently, the collective shield never returns to its initial configuration and the particles therefore never fully disentangle from the shield. 
This leads to the persistent degradation of gravitationally induced entanglement.

\begin{figure}[!th]
    \centering
    \includegraphics[width=\linewidth]{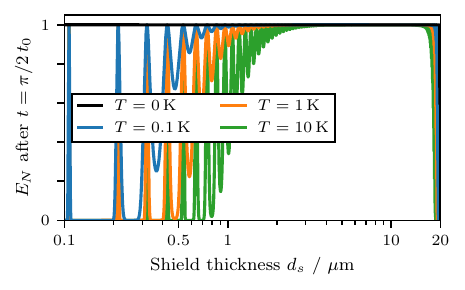}
    \caption{\small\justifying
    \textbf{GIE for different shield thicknesses.}
    Logarithmic negativity $E_N$ after an evolution time of $t=\pi/2\, t_0$ (where $E_N(T=0\si{K})=1$) in the parallel orientation for different shield thicknesses $d_s$.
    Here we show silica particles interacting with a copper shield via Casimir interactions.
    The decoherence strongly depends on the vibrational frequencies $\omega \sim d_s$, where a higher frequency increases the stability.
    For $d_s \rightarrow 2(L-R)=20\si{\mu m}$, the surface-to-surface separation between the shield and the particles becomes very small and thus Casimir interactions increase, resulting in more decoherence.}
    \label{fig:shield-thickness}
\end{figure}

We furthermore want to stress that the dynamics and entanglement suppression is strongly dependent on the size of the shield.
As $\omega_{kl}\sim d_s/r_s^2$, reducing the radius $r_s$ or increasing the thickness $d_s$ increases the eigenfrequencies and hence reduces the thermal fluctuations.
While the minimal shield radius is limited by the suppression of direct Casimir, electrostatic or magnetic couplings, the thickness can, in principle, be increased up to $d_s\sim 2(L-R)$.
The resulting entanglement after a constant time for different thicknesses $d_s$ is shown in \figref{fig:shield-thickness}.
This statement, however, only applies to noise induced by shield vibrations.
Increasing the shield thickness may instead increase noise arising from placement variations, as a thicker shield reduces surface-to-surface distance between the shield and the particles and thus increases the relevant interactions.

\subsection{Non-gravitationally mediated entanglement}
\begin{figure*}[!t]
    \centering
    \includegraphics[width=\linewidth]{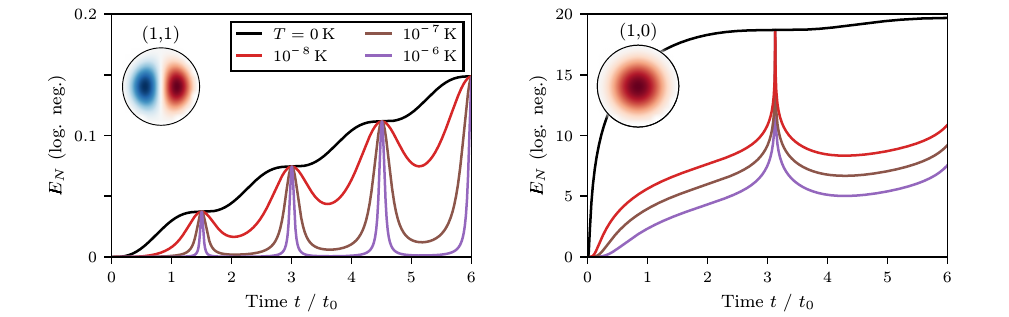}
    \caption{\justifying\small
    \textbf{Shield-mediated entanglement in the absence of direct particle--particle coupling.}
    Logarithmic negativity between two lead particles prepared in Gaussian states interacting exclusively through their common magnetic-dipole coupling to the quantum shield.
    The dynamics are shown for different shield temperatures $T$ and for the parameters from \tabref{tab:parameters}.
    \textbf{Left:} parallel orientation ($\theta=0$), for which symmetry enforces that the dominant contribution arises from the $(1,1)$ vibrational mode.
    \textbf{Right:} linear configuration ($\theta=\pi/2$), for which the strong coupling is governed primarily by the $(1,0)$-mode, whose displacement is maximal at the shields center.
    In both cases, the shield acts as an entanglement mediator, producing peaks in $E_N(t)$ of width $\sim\sqrt{\omega_{kl}/T}$ at times $t=2\pi j / \omega_{kl}$ ($j\in\mathbb{N}$).
    The black curves show the zero-temperature limit.
    For clarity, only the dominant single mode is displayed here.
    }
    \label{fig:gaussian-shield-no-gravity}
\end{figure*}
We analyze bipartite entanglement between two particles arising exclusively from their mutual Casimir or magnetic-dipole coupling to the quantum shield, while gravitational interactions are completely neglected in this section.
The dynamics are therefore fully governed by $\op H_\mathrm{Shield}$ in \Eqref{eq:hamiltonian-shield}.
The induced entanglement depends sensitively on the spatial profile $u_{kl}$ at the particles positions and on whether the particles are oriented parallel or orthogonal relative to the shield.

As discussed previously in Sec.~\ref{sec:variations}, the parallel orientation ($\theta=0$) is intrinsically robust against fluctuations in the particle--shield separation.
In this case, the dominant coupling is set by the radial gradient $\partial_r u_{kl}\big|_{r=0,\vartheta=0}$, reflecting variations in the separation distance under radial displacement.
It induces relative phases between components of a state delocalized parallel to the shield.
The lowest mode with non-zero gradient is $(1,1)$.
Its contribution leads to a weak but finite amount of shield-mediated entanglement, as shown in \figref{fig:gaussian-shield-no-gravity}.
The effect remains small because (i) the coupling is proportional to the gradient of the large Faraday shield rather than its absolute displacement at the center, and (ii) the $(1,1)$-mode possesses a comparatively high resonance frequency, which suppresses its thermal amplitude and hence its gradient.

In contrast, the linear configuration ($\theta=\pi/2$) is highly sensitive to changes in the particle--shield separation.
For a particle placed at the shield's center, the $(1,0)$-mode dominates, since its displacement attains a maximum at $r=0$.
Even at temperatures as low as $10^{-8}\si{K}$, the displacement amplitude $u_{(1,0)}(r=0)\Delta q_{(1,0)} \approx 1.4\times10^{-14}\si{m}$ exceeds the tolerance threshold on variations in the particle--shield separation established earlier in Sec.~\ref{sec:variations}.
Consequently, this yields a strong coupling between the particles and the low-frequency $(1,0)$-mode, sufficient to mediate appreciable bipartite entanglement in the absence of any direct particle--particle interaction, as shown in \figref{fig:gaussian-shield-no-gravity}.

Note, that the particle--shield coupling can get worse if the particle is not perfectly positioned in the center or if the shield is not completely homogeneous.
In such cases, the fundamental mode $(1,0)$ would also mediate small amount of additional entanglement between particles in the parallel configuration, worsen the situation discussed here.
\section{Experimental implications}\label{sec:implications}
The analysis presented allows us to identify the dominant constraints for realistic implementations of GIE experiments.
In particular, the interplay between non-gravitational interactions, geometric stability, timing precision and particle--shield couplings place strong requirements on experimentally viable platforms.

Even in the absence of interactions with the shield, slow fluctuations of the trap equilibrium position -- arising, for instance, from uncontrolled external forces -- can significantly constrain the recoverable entanglement, even when they occur only from run to run.
In particular, such fluctuations confine non-zero entanglement to narrow time windows around the trap oscillation period, whose width decreases with increasing fluctuation amplitude.
This, in turn, imposes stringent requirements on both the stability of the trapping potential and the timing precision of the measurements.

A second limitation particularly relevant for magnetically levitated systems are dominant magnetic-dipole couplings.
External trapping fields induce dipole moments in the particles, which constitute a strong non-gravitational interaction between the particles as well as between the particles and a shield.
As a consequence, magnetically levitated setups require extremely tight control of the experimental geometry, with tolerable variations in the relative placement and orientation of the detectors and the shield on the order of $\Delta L = 3\times10^{-18}\si{m}$ and $\Delta \theta = 5\times10^{-13}\si{rad}$, as seen in Figs.~\ref{fig:gauss-shield-variations}-\ref{fig:cat-detector-variations}.

In other GIE implementations, like free-fall or space-based implementations, where trapping potentials are absent during the entanglement generation step, the dominant non-gravitational couplings arise from Casimir interactions of the particles with the shield or from residual electrostatic or magnetostatic fields.
Geometric stability remains a major challenge, with tolerable variations of at least $\Delta L = 4\times10^{-15}\si{m}$ and $\Delta\theta = 10^{-9}\si{rad}$ required in order to preserve the gravitationally generated entanglement signal (see Figs.~\ref{fig:gauss-shield-variations}-\ref{fig:cat-detector-variations}).
These variations strongly depend on the orientation of the principal delocalization axes of the particles and can be estimated analytically by \Eqref{eq:log-neg-cat-variations}.
We find that the parallel configuration, in which the principal delocalization axis lies parallel to the shield is considerably more stable against variations in the position.
The linear configuration, where particles are oriented perpendicular to the shield, exhibits advanced stability against orientation noise.
The choice of orientation therefore provides an additional degree of freedom for optimizing experimental robustness.

We also find that the orientation of the experimental apparatus itself has to account for the change of the effective direction of the gravitational pull of the Sun and other celestial bodies, which varies due to Earth’s rotation.
It is worth noting that, in all cases considered here, independent measurements of the relative positions between all relevant objects can be used to mitigate decoherence in post-processing.
Additionally, even a perfectly static configuration can lead to decoherence if there is uncertainty in the duration of experimental evolution times.

In Sec.~\ref{sec:shield}, we further investigated the role of the shield whose vibrational degrees of freedom are treated dynamically.
Thermally populated vibrational modes were shown to induce decoherence via entanglement with the particles and, simultaneously, to mediate non-gravitationally induced entanglement between them, thereby potentially obscuring the gravitational signal.
The resulting decoherence rates depend sensitively on the shield's geometry and size:
reducing the lateral extent of the shield or increasing its thickness, substantially suppresses the coupling between vibrational modes and the particles.
Modifying the shield geometry to a non-planar shield (like a concave parabolic or hollow half-spherical shield that is thin at its center and thicker towards the edges) can simultaneously reduce static Casimir forces (as suggested by Refs.~\cite{Zaheer_2010, Graham_2011}) and suppress vibrational amplitudes further.
\section{Conclusion}\label{sec:conclusion}
In this work, we have analyzed the impact of geometric fluctuations of the experimental setup in experiments aimed to observe gravitationally induced entanglement (GIE) between spatially delocalized masses.
We considered run-to-run fluctuations in the particle preparation and trapping, detector positioning and shield placement.
Our results show that a high degree of control over the placement and orientation of all relevant parts is required.
Most notably, due to strong requirements on the stability of the trapping potential and measurement timing, GIE implementations with magnetically levitated nanoparticles appear to be particularly challenging.
By treating both, the particles and the vibrational degrees of freedom of the shield within a fully quantum framework, we have shown that Casimir and magnetic-dipole interactions impose strict limitations in realistic experimental settings.

Beyond the geometric variations considered here, particles near material surfaces are also subject to surface-induced noise from electromagnetic field fluctuations, as described within macroscopic dielectric-response frameworks~\cite{Martinetz2022}.
Such approaches account for coherent surface interactions as well as temperature-dependent electromagnetic fluctuations of the surface that lead to heating and decoherence, even for a perfectly fixed shield.
By contrast, the mechanism analyzed in this work arises from stochastic geometric variations of the shielding element, which modulate coherent particle--shield interactions from run to run.
The resulting degradation of entanglement is therefore distinct in origin from intrinsic electromagnetic surface noise and does not rely on thermal fluctuations.
Our framework enables us to derive quantitative bounds on the positioning and alignment stability required of the shield in order to resolve gravitationally mediated entanglement.

Taken together, our results establish quantitative constraints on placement precision, angular stability, measurement timing, temperature, and shield geometry that must be satisfied in order to observe gravitationally induced entanglement in the presence of realistic electromagnetic shielding.
Beyond existing works~\cite{Kamp2020,Schut2023}, we present a setup-independent analysis of Casimir- and magnetic-dipole induced noise arising from both stochastic preparation and measurement errors and thermally excited shield vibrations.
We explicitly compare the resulting decoherence and entanglement suppression for initially prepared Schrödinger-cat states and Gaussian squeezed states -- cases that, to our knowledge, have not previously been examined in this context.
We find that at small times, entanglement generation and required placement stability does not appreciably depend on the class of state, as long as the position variance $(\Delta x)^2$ is fixed.
By treating the vibrational degrees of freedom of the Faraday shield as a fully quantum multi-mode system rather than a static boundary, our framework consistently captures both decohering and non-gravitationally-mediated entangling effects, providing a unified description of non-gravitational limitations in GIE experiments.

\medskip
\paragraph*{Acknowledgments.} This work was supported by the ERC Synergy Grant HyperQ (grant no. 856432) and the DFG via QuantERA project Lemaqume (grant no 500314265).

\bibliography{bib}

\appendix
\onecolumngrid
\section{Size and thickness of the shield}\label{apx:shield-size}
\subsection{Thickness estimations}
The shield is used to screen direct non-gravitational interactions between the two particles.
It can either be a conductive Faraday shield to screen electrostatic- and Casimir interactions, which impose dominant noise in GIE experiments with silica nanospheres or a superconducting Meissner shield for dominant magnetic-dipole interactions in GIE experiments with magnetically levitated superconducting lead particles.

We focus mainly on Coulomb interactions (if both particles carry a net charge $q_{A/B}$), electric and magnetic dipole-dipole interactions (either through intrinsic dipole moments or externally induced by stray fields, patch potentials or the trapping mechanism) and mutual Casimir interactions.
The transmission of the electromagnetic field through a conductor of thickness $d_s$ and conductivity $\sigma$ (with $\sigma_\mathrm{Copper}\approx 1.5\times 10^{10}\si{S/m}$ at low temperatures~\cite{Berman1952}) is approximated by $\mathcal{T}=2/(Z_0 \sigma d_s)$, where $Z_0 = 377\si{\Omega}$ is the impedance of free space~\cite{Vandenbosch2022}.
Demanding that the entanglement rate $\Gamma_\mathrm{ent.} = \dd E_N / \dd t \big|_{t=0} = 2 \Delta x_A\Delta x_B\norm{\partial^2_r V\big|_{r=2L}} / (\hbar \log2)$ due to gravitational interactions (see \Eqref{eq:gaussian-entanglement-rate} or Ref.~\cite{Pedernales2023}) is strictly larger (ideally by a constant factor) than other direct interactions $V(r)$, the thickness of the shield can be estimated.

The corresponding entanglement rates for each interaction are given by:
\begin{align}
    \Gamma_\mathrm{Gravity} &= \frac{G}{2\hbar \log2}\frac{M_A M_B \Delta x_A \Delta x_B}{L^3} \\
    \Gamma_\mathrm{Casimir} &= \mathcal{T}^2 \frac{161 c}{128\pi \log2}\frac{R^6 \Delta x_A \Delta x_B}{ L^9}\left(\frac{\veps_r - 1}{\veps_r + 2}\right)^2
    \quad \text{for} \quad V(r) = -\frac{23 \hbar c R^6}{4\pi r^7}\left(\frac{\veps_r - 1}{\veps_r + 2}\right)^2,\ \text{from Ref.}~\cite{Emig2007} \\
    \Gamma_\mathrm{Coulomb} &= \mathcal{T}\frac{1}{8\pi\veps_0 \hbar \log2}\frac{\abs{q_Aq_B}\Delta x_A\Delta x_B}{L^3} 
    \qquad \text{for} \quad V(r) = \frac{1}{4\pi\veps_0}\frac{q_Aq_B}{r} \\
    \Gamma_\mathrm{elec.\ Dipole} &= \mathcal{T}^2 \frac{3}{8\pi\veps_0 \hbar \log2}\frac{\abs{p_Ap_B}\Delta x_A\Delta x_B}{L^5}
    \qquad \text{for} \quad V(r) = -\frac{2}{4\pi\veps_0}\frac{p_Ap_B}{r^3}, \text{ (linear dipole alignment)}
\end{align}
The entanglement rate is suppressed by a factor of $\mathcal{T}$ for Coulomb interactions and by a factor of $\mathcal{T}^2$ for dipole-dipole and Casimir interactions, as their coupling arises from field correlations that require two transmissions through the shield (from $B\rightarrow A$ and back $A\rightarrow B$).
The resulting estimated minimum thickness of the shield for the different interactions are summarized in \tabref{tab:shield-size}.
\begin{table*}[!th]
    \caption{\justifying\small
    Size and thickness estimations of the shield (either conductive Faraday- or superconducting Meissner shielding) to effectively shield different forms of direct particle-particle interactions below the gravitational coupling strength.
    Optical skin depth ($\sim 20\si{nm}$~\cite{Roberts1960}), mechanical rigidity and ease of fabrication ultimately impose realistic lower bounds on the shields thickness of around $1-10\si{\mu m}$.}
    \label{tab:shield-size}
    \begin{ruledtabular}
    \begin{tabular}{l l l c c}
    Interaction & Note & type of shielding & min. thickness $d_s$ & min. radius $r_s$ \footnote{The required radius is estimated below in Sec.~\ref{sec:apx:lateral-shield-size}.} \\
    \hline
    \textbf{Casimir} & & cond. Faraday & $3\times 10^{-10}\si{m}$ & $50\si{\mu m}$ \vspace{4pt} \\ 
    \multirow{2}{*}{\textbf{mag. Dipole}} & Silica: $B_\mathrm{ext.}\lesssim 71\si{\mu T}$ must hold & none / mu-metal & & \\
    & Supercond. lead: $B_\mathrm{ext.}\lesssim 4.25\si{nT}$ must hold & supercond. & $440\si{nm}$ \footnote{Estimated using the London penetration depth of the superconducting material in \Eqref{eq:apx:thickness-superconducting-shield}.} & $7.7\si{mm}$ \vspace{4pt} \\
    \textbf{Coulomb} & Each particle charged with $q_{A/B}=e$ & cond. Faraday & $2 \times 10^{-8}\si{m}$ & $66\si{cm}$ \vspace{4pt} \\
    \multirow{2}{*}{\textbf{elec. Dipole}} & Intrinsic dipole moment $p\approx 10^{-2}e\si{cm}$ & \multirow{2}{*}{cond. Faraday} & $5\times 10^{-10}\si{m}$ & $1.2\si{mm}$ \\
    & Induced dipole (e.g. patch potentials) $p\lesssim 10^{-3} e\si{cm}$ & & $5\times 10^{-11}\si{m}$ & $180\si{\mu m}$
    \end{tabular}
    \end{ruledtabular}
\end{table*}
One has to note, that these values are very thin.
The requirement, that the shield has to act as a perfect mirror for the electromagnetic modes over all relevant frequencies (around $\lambda\sim 2L$, see e.g. chapter 12 in Ref.~\cite{Bordag2009} for the relevance to Casimir interactions) is much more stringent.
The optical skin depth at these frequencies is usually around $20\si{nm}$~\cite{Roberts1960} and facilitates a more realistic lower bound on the thickness.
The thickness for a conducting Faraday shield is therefore mainly dictated by mechanical rigidity and ease of fabrication rather than by fundamental physical limits.

Static magnetic fields are usually shielded very poorly by a Faraday shield and require either Mu-metal shields or, as focused on here, superconducting Meissner shielding.
For an induced magnetic-dipole with $\vec{m}=4\pi R^3 \chi_V \vec{B} / (3 \mu_0)$ ($\chi_V$ being the volume susceptibility with $\chi_{V,\,\text{SiO}_2}\approx -1.4\times 10^{-5}$~\cite{CRC2017} and $\chi_{V,\,\mathrm{supercond.}}=-1$), the entanglement rate between both particles in the absence of magnetic shielding is given by
\begin{equation}
    \Gamma_\mathrm{mag.\ Dipole} = \frac{3\mu_0}{8\pi \hbar \log2} \frac{\abs{m_A m_B}\Delta x_A \Delta x_B}{L^5} 
    \qquad \text{for} \quad V(r) = -\frac{2\mu_0}{4\pi}\frac{m_Am_B}{r^3}, \text{ for linear dipole alignment} .
\end{equation}
Estimating the upper limit of external magnetic fields before these dipole interactions exceed gravitational entanglement generation, we find $B \lesssim 71\si{\mu T}$ for silica particles and $B\lesssim 4.25 \si{nT}$ for superconducting lead.
A superconducting Niobium shield with London penetration depth $\lambda_L\approx 40\si{nm}$~\cite{Maxfield1965} would suppress a trap field of $B_\mathrm{trap}\sim 250\si{\mu T}$ to the allowed values for a thickness of
\begin{equation}\label{eq:apx:thickness-superconducting-shield}
    d_s \geq \lambda_L \log(\frac{B_\mathrm{trap}}{B}) \approx 440\si{nm}.
\end{equation}

\subsection{Lateral size estimations}\label{sec:apx:lateral-shield-size}
The lateral dimension $r_s$ of the shield can be estimated by considering fringe fields and diffraction around the shield.
If a fraction $\kappa\in[0,1]$ of the electromagnetic modes can leak around the edges of the shield and re-establish a direct non-gravitational coupling.
Geometric considerations yield that for a potential $V\sim L^{-d}$ ($d=1$ for the Coulomb potential, $d=3$ for dipole-dipole coupling and $d=7$ for Casimir interactions), a shield of radius
\begin{equation}
    r_s \geq L\tan(\arccos(\kappa^{1/d})) = L\sqrt{\kappa^{-2/d} - 1}
\end{equation}
is required.
Since the entanglement rate is proportional to the underlying interaction potential, one can simply demand $\abs{V_\mathrm{Gravity}} > \kappa \abs{V_\mathrm{non-gravitational}}$ to estimate the radius of the shield such that the gravitational interaction dominates.
The resulting values for the different interactions are summarized in \tabref{tab:shield-size}.

\section{Gravitational and non-gravitational interactions}\label{apx:interactions}
\subsection{Gravitational interaction}\label{apx:interactions-gravity}
The gravitational interaction between two particles is given by
\begin{equation}\label{eq:apx:gravity-hamiltonian}
    \hat{H}_\mathrm{Gravity} = -\frac{G M_A M_B}{\big|\hat{d}_{AB}\big|},
\end{equation}
where $G$ is the gravitational constant, $M_{A/B}$ are the masses of the two particles, and $\big|\hat{d}_{AB}\big|$ is the canonically quantized distance between them, defined as
\begin{equation}
    \hat{d}_{AB} = \sqrt{(\hat{x}'_A - \hat{x}'_B)^2 + (\hat{y}'_A - \hat{y}'_B)^2} .
\end{equation}
Here, $\hat{x}'_{A/B}$ and $\hat{y}'_{A/B}$ are the absolute position operators in the lab frame of the respective particles.
As the particles' wave functions are centered around $(x_{A/B}'=\pm L, y_{A/B}' = 0)$ and their principal axes of delocalization are tilted by an angle $\theta_{A/B}$ with respect to shield, the particles positions can be expressed in their local position operators $\op x_{A/B},\,\op y_{A/B}$ as
\begin{equation}
    \begin{pmatrix} 
        \hat{x}'_{A/B} \\[4pt] 
        \hat{y}'_{A/B} 
    \end{pmatrix}
    =
    \begin{pmatrix} 
         \pm L + \hat{x}_{A/B} \sin\theta_{A/B} - \hat{y}_{A/B} \cos\theta_{A/B} \\[4pt]
         \hat{x}_{A/B} \cos\theta_{A/B} + \hat{y}_{A/B} \sin\theta_{A/B}
    \end{pmatrix}.
\end{equation}
Since $\Delta x, \Delta y \ll L$, $\hat{H}_\mathrm{Gravity}$ can be expanded up to second order in $\hat{x}$ and $\hat{y}$. 
For simplicity, we restrict ourselves to the parallel configuration with $\theta_{A/B} = 0$ and the linear configuration with $\theta_{A/B} = \pi/2$, obtaining
\begin{align}\label{eq:apx:gravity-parallel}
    \hat{H}_\mathrm{Gravity,\, parall.} 
    &\approx \lambda \Big( 
        - L \hat{y}_A - \tfrac{1}{2}\hat{y}_A^2 
        + L \hat{y}_B - \tfrac{1}{2}\hat{y}_B^2 
        + \hat{y}_A \hat{y}_B   
        + \tfrac{1}{4}\hat{x}_A^2 + \tfrac{1}{4}\hat{x}_B^2 
        - \tfrac{1}{2}\hat{x}_A \hat{x}_B 
    \Big), \\
    \hat{H}_\mathrm{Gravity,\, linear} 
    &\approx \lambda \Big( 
        L \hat{x}_A - \tfrac{1}{2}\hat{x}_A^2 
        - L \hat{x}_B - \tfrac{1}{2}\hat{x}_B^2 
        + \hat{x}_A \hat{x}_B    
        + \tfrac{1}{4}\hat{y}_A^2 + \tfrac{1}{4}\hat{y}_B^2 
        - \tfrac{1}{2}\hat{y}_A \hat{y}_B 
    \Big). \label{eq:apx:gravity-linear}
\end{align}
Here, the parameter
\begin{equation}\label{eq:apx:gravitational-coupling}
    \lambda = \frac{G M_A M_B}{4 L^3},
\end{equation}
with units $\mathrm{J/m^2}$ has been introduced. 
Entanglement is generated by the terms coupling the position of particle $A$ with particle $B$, i.e. terms proportional to $\op{x}_A\op{x}_B$ and $\op{y}_A\op{y}_B$.

For simplicity, we neglect variations in $L$ and $\theta$ in the gravitational interaction. 
The associated gravitational fluctuations are suppressed by $\lambda/\eta \sim 10^{-7}$ relative to the Casimir coupling $\eta$ for silica particles (defined later in \Eqref{eq:apx:casimir-coupling}) and suppressed by a factor of $\lambda/\delta \sim 10^{-11}$ relative to magnetic-dipole interactions in superconducting lead particles ($\delta$ defined in \Eqref{eq:apx:mag-dipole-coupling}).
Furthermore, numerical simulations confirm that entanglement degradation is driven almost entirely by non-gravitational interactions.

\subsection{Casimir interactions}\label{apx:interactions-casimir}
Closed-form expressions for this potential across all distances are not known, however approximations in the two limiting regimes for small and large distances exists.
For small surface-to-surface separations $L-R-\tfrac{d_s}{2} \ll R$, the interaction can be described by the proximity-force-approximation (PFA)~\cite{Bulgac2006, Pirozhenko2013}
\begin{equation}
    V_\mathrm{PFA} = -\frac{\hbar c\pi^3 R}{720(L-R-d_s/2)^2} \vphi(\veps_r)
\end{equation}
where $L$ is the distance from the sphere's center to the plane, $\hbar$ is planks constant, $c$ the speed of light and $\vphi(\veps_r)$ accounts for dielectric properties. 
It is given by $\vphi(\veps_r)=(\veps_r-1)/(\veps_r+1)\chi(\veps_r)$ with $\chi(\veps_r \rightarrow\infty)\rightarrow 1$, $\chi(\veps_r = 1) \approx 0.46$ and $\chi(\veps_{r,\mathrm{Silica}})\approx 0.5$ being a tabulated function~\cite{Lifshitz1992}.
In the large separation limit (LSL) $L\gg R$, the interaction reduces to the well-known Casimir-Polder potential~\cite{CasimirPolder1948} in leading order~\cite{Pirozhenko2013,Emig2008}, i.e.
\begin{equation}\label{eq:casimir-lsl}
    V_\mathrm{LSL}=-\frac{3\hbar c}{8\pi L^4}\alpha \quad \text{with } \alpha= \left(\frac{\veps_r - 1}{\veps_r + 2}\right) R^3\, ,
\end{equation}
where $\alpha$ is the electric polarizability of a sphere. 
As shown in Figure 7 of Ref.~\cite{Emig2008}, the PFA consistently predicts stronger interactions for a sphere-plate geometry and thus provides a conservative worst-case estimate of their influence throughout the whole range of $L/R\in[1,\infty)$.
We therefore use the PFA in all subsequent calculations, as it provides an upper bound on the noise induced on the particles.
For the parameters in \tabref{tab:parameters} (i.e. $R/L=0.5$), the Casimir energy is overestimated roughly by a factor of $2.4$.
\begin{figure}[!thb]
    \centering
    \includegraphics[width=\linewidth]{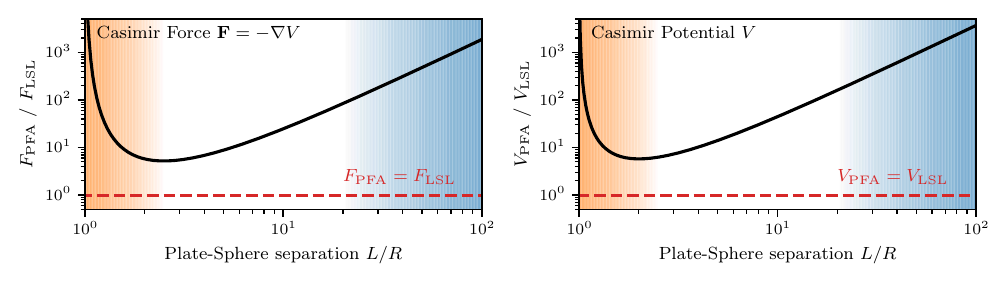}
    \caption{\small\justifying
    \textbf{Comparison of Casimir interaction models.} 
    The Proximity-Force Approximation (PFA, validity region shaded orange), valid in short-distance regimes $L\lesssim 2R$, consistently predicts a larger Casimir potential and attraction force $\vec{F}=-\nabla V$ than the large-separation limit (LSL, shaded blue), which applies for $L \gg R$.
    Here, $L$ is the center-surface separation and $R$ the spheres radius. 
    Because the PFA exceeds the LSL for all ratios $L/R$, it provides a conservative upper bound on the Casimir interaction strength and is thus used throughout this work.
    }
    \label{fig:apx:casimir-comparison}
\end{figure}

The corresponding Hamiltonian between one particle and the shield is given by
\begin{equation}\label{eq:apx:casimir-hamiltonian}
    \op{H}_\mathrm{Casimir}=-\frac{\hbar c \pi^3 R}{720}\vphi(\veps_r) \frac{1}{|\hat{d}|^2}
\end{equation}
where $\hat{d}$ is the surface-to-surface separation between particle and shield. Including variations in $L$ and $\theta$, it is given by
\begin{equation}
    \hat{d} = d_0 + \xi^{(L)}_{A/B} \mp \left[\op{x}\sin(\theta_{A/B} + \xi^{(\theta)}_{A/B}) - \op{y}\cos(\theta_{A/B} + \xi^{(\theta)}_{A/B}) \right]
\end{equation}
where the \q{$\,\pm\,$} distinguishes between particle $A$ and $B$ and $d_0 =L - R -\tfrac{d_s}{2}$.
Expanding \Eqref{eq:apx:casimir-hamiltonian} in $\op{x},\op{y}$ (since $\Delta x, \Delta y \ll L-R-d_s/2$) as well as in the variations $\xi^{(L)}\ll L-R-d_s/2$ and $\xi^{(\theta)}\ll (L-R-d_s/2)/\Delta x$, we find
\begin{multline}\label{eq:apx:Casimir-parallel}
    \op H_\mathrm{Casimir,\,parall.} \approx \eta \Big[
        \pm 2d_0\op y - 3\op y^2 
        \mp 6\xi^{(L)}\op y + \frac{12\xi^{(L)}\op y^2}{d_0} \pm \frac{12 \xi^{(L)\,2} \op y}{d_0} - \frac{30 \xi^{(L)\,2} \op y^2}{d_0^2} \\
        \mp 2d_0 \xi^{(\theta)} \op x + 6 \xi^{(\theta)}\op x \op y - 3\xi^{(\theta)\,2} \op x^2 \mp d_0 \xi^{(\theta)\,2} \op y + 3\xi^{(\theta)\,2} \op y^2
    \Big]
\end{multline}
\begin{multline}\label{eq:apx:Casimir-linear}
    \op H_\mathrm{Casimir,\,linear.} \approx \eta \Big[
        \mp 2d_0\op x - 3\op x^2 
        \pm 6\xi^{(L)}\op x + \frac{12\xi^{(L)}\op x^2}{d_0} \mp \frac{12 \xi^{(L)\,2} \op x}{d_0} - \frac{30 \xi^{(L)\,2} \op x^2}{d_0^2} \\
    \mp 2d_0 \xi^{(\theta)} \op y - 6 \xi^{(\theta)}\op x \op y + 3\xi^{(\theta)\,2} \op x^2 \pm d_0 \xi^{(\theta)\,2} \op x - 3\xi^{(\theta)\,2} \op y^2
    \Big]
\end{multline}
As before, we restricted ours elves to the linear and parallel case only and introduced an parameter $\eta$ with dimensions of energy over length squared which is given by
\begin{equation}\label{eq:apx:casimir-coupling}
    \eta = \frac{\hbar c \pi^3 R}{720 (L-R-d_s/2)^4}\vphi(\veps_r) .
\end{equation}

\subsection{Magnetic-dipole interactions}\label{apx:interactions-mag-dipole}
The interaction Hamiltonian between a sphere of radius $R$ with magnetic volume susceptibility $\chi_V$ in the presence of an external field $\vec{B}_\mathrm{ext.}$ with a superconducting shield is given by
\begin{equation}\label{eq:apx:mag-dipole-hamiltonian}
    \op H_\mathrm{mag.\, Dipole} = -\frac{\abs{\vec{m}}^2\mu_0}{32\pi |\op d|^3}\left(1+\cos^2\phi\right) 
\end{equation}
with the magnetic-dipole moment $\vec{m}=4\pi R^3 \chi_V \vec{B}_\mathrm{ext.}/(3\mu_0)$ and vacuum permeability $\mu_0$.
The angle $\phi$ is measured between the dipole orientation and the shield's normal.
We take $\phi=0$ for a worst-case estimate. 
The shield-particle separation $\op d$ is given by
\begin{equation}
    \op d = d_0 \pm \xi^{(L)} \mp \left[\op x \sin(\theta_{A/B} + \xi^{(\theta)}) - \op y \cos(\theta_{A/B} + \xi^{(\theta)})\right]
\end{equation}
where the $\pm$ distinguishes between particle $A$ and $B$ and $d_0 = L-\tfrac{d_s}{2}$.
As done previously for Casimir interactions, we can expand the Hamiltonian in second order in $x_{A/B}$, $y_{A/B}$ and in the variations $\xi^{(L)},\xi^{(\theta)}$ for the parallel and linear orientation:
\begin{multline}\label{eq:apx:mag-dipole-parallel}
    \op H_\mathrm{mag.\,Dipole,\,parall.} \approx \delta \Big[
        \pm 3d_0\op y - 6\op y^2 
        \mp 12\xi^{(L)}\op y + \frac{30\xi^{(L)}\op y^2}{d_0} \pm \frac{30 \xi^{(L)\,2} \op y}{d_0} - \frac{90 \xi^{(L)\,2} \op y^2}{d_0^2} \\
        \mp 3d_0 \xi^{(\theta)} \op x + 12 \xi^{(\theta)}\op x \op y - 6\xi^{(\theta)\,2} \op x^2 \mp \frac{3}{2}d_0 \xi^{(\theta)\,2} \op y + 6\xi^{(\theta)\,2} \op y^2
    \Big]
\end{multline}
\begin{multline}\label{eq:apx:mag-dipole-linear}
    \op H_\mathrm{mag.\,Dipole,\ linear} \approx \delta \Big[
        \mp 3d_0\op x - 6\op x^2 
        \pm 12\xi^{(L)}\op x + \frac{30\xi^{(L)}\op x^2}{d_0} \mp \frac{30 \xi^{(L)\,2} \op x}{d_0} - \frac{90 \xi^{(L)\,2} \op x^2}{d_0^2} \\
        \mp 3d_0 \xi^{(\theta)} \op y - 12 \xi^{(\theta)}\op x \op y + 6\xi^{(\theta)\,2} \op x^2 \pm \frac{3}{2} d_0 \xi^{(\theta)\,2} \op x - 6\xi^{(\theta)\,2} \op y^2
    \Big]
\end{multline}
Here, we defined the coupling strength with units $\mathrm{J/m^2}$ as
\begin{equation}\label{eq:apx:mag-dipole-coupling}
    \delta = \frac{2\abs{\vec{m}}^2 \mu_0}{32\pi(L-d_s/2)^5} .
\end{equation}

\subsection{Dynamical shield}\label{apx:interactions-shield}
The shapes of the vibrational eigenmodes, labeled by integers $(k,l)$ with $k \in [1,\infty)$ and $l \in [0,\infty)$, can be expressed as~\cite{Rao2019}
\begin{equation}
    u_{kl}(r,\vartheta,t) =\left[J_l(\beta_k r) - \frac{J_l(\beta_k r_s)}{I_l(\beta_k r_s)} I_l(\beta_k r) \right]
    \cos(l\vartheta) \sin(\omega_{kl} t),
    \label{apx:eq:ukl}
\end{equation}
where $\beta_k := \tilde{r}_k / r_s$, and the eigenfrequencies are given by
\begin{equation}
    \omega_{kl}
    = \tilde{r}_k^2 \frac{d_s}{r_s^2} 
    \sqrt{\frac{E}{12 \rho (1 - \nu^2)}}.
\end{equation}
\begin{figure}[!th]
    \centering
    \includegraphics[width=\linewidth]{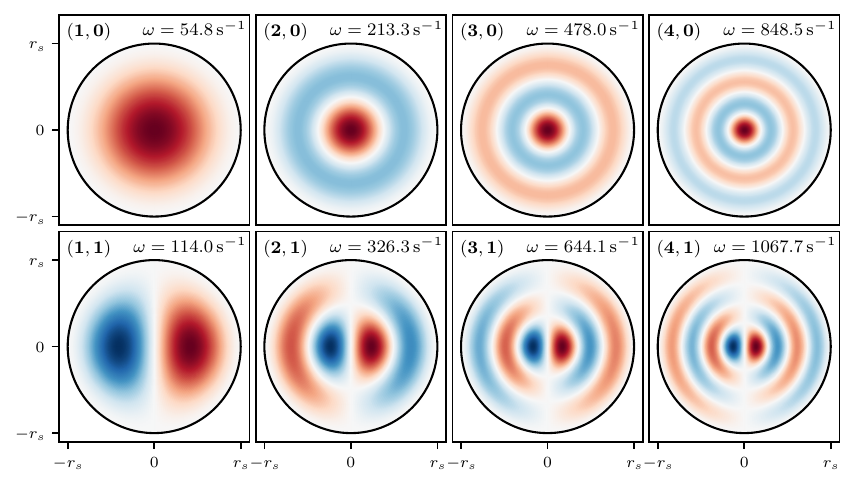}
    \caption{\small\justifying
    \textbf{Selected vibrational modes of a circular shield.}
    Spatial profiles $u_{kl}(r,\vartheta)$ of the lowest vibrational modes, together with their corresponding eigenfrequencies, computed for a Copper shield with the parameters given in \tabref{tab:parameters}.
    For a conservative estimation of shield-induced decoherence, the particle is assumed to be located at the center of the shield ($r=0$), where the displacements of the $(k,0)$-modes are maximal and where the radial gradients $\partial_r u_{kl}\big|_{r=0}$ of the $(k,1)$-modes attains their largest value for $\vartheta=0$.
    }
    \label{fig:shield-modes}
\end{figure}
Here, $E$ is the Young’s modulus (with $E_\mathrm{Cu}=110\si{GPa}$ for a Copper shield and $E_\mathrm{Nb}=105\si{GPa}$ for Niobium), $\nu$ the Poisson ratio ($\nu_\mathrm{Cu}=1/3$ and $\nu_\mathrm{Nb}=0.40$), and $\rho$ the mass density of the shield’s material ($\rho_\mathrm{Cu}=8960\si{kg/m^3}$ and $\rho_\mathrm{Nb} = 8582 \si{kg/m^3}$).
The quantity $\tilde{r}_k$ denotes the $k$-th root of the characteristic equation 
\begin{equation}
    J_l(\tilde{r}_k) I_{l+1}(\tilde{r}_k) + I_l(\tilde{r}_k) J_{l+1}(\tilde{r}_k) = 0,
\end{equation}
with $J_l$ and $I_l$ being the Bessel- and modified Bessel functions of the first kind, respectively.
For a worst-case estimate, we set $r=0$ and $\vartheta=0$ (see \figref{fig:shield-modes} and main text).
If the spatial extend of the superposition is small compared to the scale over which the mode-profile varies (i.e. if $\Delta x \partial_r u_{kl},\ \Delta y \partial_r u_{kl} \ll 1$), the mode can be linearized around the particles' delocalization, spanning $\op{r} = \left(\op{x}\cos\theta - \op{y}\sin\theta\right)$, as
\begin{equation}
    u_{kl}(\op{r}) \approx u_{kl}(0) + \op{r} \partial_r u_{kl} \big|_{r=0} .
\end{equation}
The distance between the particle and the shield, $\hat{d}_{kl}$, depends on $u_{kl}$ and is thus given by
\begin{equation}
    \hat{d}_{kl} = d_0
    \pm (\hat{x} \sin \theta_{A/B} + \hat{y} \cos \theta_{A/B}) 
    \mp \hat{q}_{kl} \Big[ u_{kl}(0) \mp \left(\op{x}\cos\theta_{A/B} - \op{y}\sin\theta_{A/B}\right)\partial_r u_{kl}\big|_{r=0} \Big],
\end{equation}
where $\hat{q}_{kl}$ is the shield's position operator.
For Casimir interactions between silica particles, we find $d_0 = L - R -\frac{d_s}{2}$. For $\Delta x, \Delta y \ll d_0$, the Hamiltonian \Eqref{eq:apx:casimir-hamiltonian} can be expanded in $\op{x}$, $\op{y}$ and $\op{q}_{kl}$ to yield
\begin{align}\label{eq:hamiltonian-particle-shield-parallel}
    \hat{H}_{kl, \mathrm{Casimir,\, parall.}}
    &\approx \eta \Big[
        \mp 2 d_0 u_{kl}(0) \hat{q}_{kl}
        \pm 2 d_0 \hat{y}
        - 3 u_{kl}(0)^2 \hat{q}_{kl}^2
        + 6 u_{kl}(0) \hat{q}_{kl} \hat{y}
        - 3 \hat{y}^2
        + 2 d_0 \partial_r u_{kl}\big|_{r=0} \hat{q}_{kl} \hat{x}
    \Big],\\
    \label{eq:hamiltonian-particle-shield-linear}
    \hat{H}_{kl, \mathrm{Casimir,\, linear}}
    &\approx \eta \Big[
        \mp 2 d_0 u_{kl}(0) \hat{q}_{kl}
        \pm 2 d_0 \hat{x}
        - 3 u_{kl}(0)^2 \hat{q}_{kl}^2
        + 6 u_{kl}(0) \hat{q}_{kl} \hat{x} 
        - 3 \hat{x}^2
        - 2 d_0 \partial_r u_{kl}\big|_{r=0} \hat{q}_{kl} \hat{y}
    \Big] .
\end{align}

For magnetic-dipole interactions between superconducting lead particles, $d_0$ is simply given by $d_0=L - d_s/2$.
Similar to above, the Hamiltonian \Eqref{eq:apx:mag-dipole-hamiltonian} can be expanded like
\begin{align}
    \op H_{kl, \mathrm{mag.\,Dipole,\, parall}} &
    \approx \delta \Big[
        \mp 3d_0u_{kl}(0) \op q_{kl} 
        \pm 3d_0 \op y 
        - 6 u_{kl}(0)^2 \op q^2_{kl} 
        + 12 u_{kl}(0) \op q_{kl} \op y
        - 6\hat y^2
        + 3d_0 \partial_ru_{kl}\big|_{r=0} \op q_{kl} \op x 
    \Big]\, ,
    \\
    \op H_{kl, \mathrm{mag.\,Dipole,\, linear}} &
    \approx \delta \Big[
        \mp 3d_0u_{kl}(0) \op q_{kl} 
        \pm 3d_0 \op x
        - 6 u_{kl}(0)^2 \op q^2_{kl} 
        + 12 u_{kl}(0) \op q_{kl} \op x
        - 6\hat x^2
        - 3d_0 \partial_ru_{kl}\big|_{r=0} \op q_{kl} \op y
    \Big] .
\end{align}
\section{Gaussian states}
\subsection{Time evolution of the covariance matrix}\label{apx:time-evolution-gaussian-formalism}
For a system described by the quadrature operators $\mathbf{r} = (x_1, p_1, \dots, x_N, p_N)^\intercal$ obeying the canonical commutation relations
\begin{equation}
    [\mathbf{r}, \mathbf{r}^\intercal] = i\hbar\Omega, \qquad
    \Omega := \begin{pmatrix}
        0 & 1 \\ -1 & 0
    \end{pmatrix}^{\oplus N},
\end{equation}
the statistical properties of a Gaussian state are fully determined by its first and second moments
\begin{equation}
    d_i = \mean{r_i}, \quad V_{ij} = \frac{1}{2}\mean{r_i r_j + r_j r_i} ,
\end{equation}
where $\mean{\,\cdot\,}=\tr{\,\cdot\,\rho}$ denotes the expectation value.
Gaussianity is preserved under a time evolution of a quadratic Hamiltonian in the form
\begin{equation}
    \hat{H} = \tfrac{1}{2}\mathbf{r}^\intercal G \mathbf{r} + \mathbf{g}^\intercal\mathbf{r} .
\end{equation}
and linear Lindblad operators
\begin{equation}
    L_k = \mathbf{l}_k^\intercal \mathbf{r},\quad\mathbf{l}_k\in\mathbb{C}^{2N}
    \quad\text{with Kossakowski matrix}\quad
    C\equiv\sum_k \mathbf{l}_k\mathbf{l}_k^\intercal .
\end{equation}
The (adjoint) Heisenberg generator of the dynamics for an operator $\op O$ is then then given by
\begin{equation}
    \dv{t} \op{O} = \frac{i}{\hbar}[\op H,\op O] + \sum_k\left(L_k^\dagger \op O L_k - \frac{1}{2}\left\{ L_K^\dagger L_K,\op O\right\}\right) .
\end{equation}
This results in the following dynamics of $\mathbf{r}$ and $\mathbf{rr}^\intercal$:
\begin{align}
    \dot{\mathbf{r}} &= K \mathbf{r} + \Omega \mathbf{g},
    \qquad \text{with} \qquad
    K = \Omega G - \Omega \Im(C) \\
    \dv{t}(\mathbf{rr}^\intercal) &= \mathbf{\dot{r} r}^\intercal + \mathbf{r \dot{r}}^\intercal + \mathcal{D}
    \qquad \text{with}\qquad
    \mathcal{D} = \hbar^2\Omega \Re(C) \Omega^\intercal .
\end{align}
The dynamics for $\mathbf{d}$ and $V$ follow directly as
\begin{equation}
    \dot{\mathbf{d}} = K\mathbf{d} + \Omega\mathbf{g} 
    \qquad\text{with solution}\qquad
    \mathbf{d}(t) = S(t) \mathbf{d}(0) + \mathbf{h}(t)
\end{equation}
and
\begin{equation}
    \dot{V} = KV + V K^\intercal + \Omega\mathbf{g}\mathbf{d}^\intercal + \mathbf{d}(\Omega\mathbf{g})^\intercal + \mathcal{D}
\end{equation}
which, in general, is solved by
\begin{equation}
    V(t) = S(t)V(0)S(t)^\intercal + S(t)\mathbf{d}(0)\mathbf{h}(t)^\intercal + \mathbf{h}(t)(S(t)\mathbf{d}(0))^\intercal + \mathbf{h}(t)\mathbf{h}(t)^\intercal + \int_0^t\dd s\,S(t-s) \mathcal{D} S(t-s)^\intercal .
\end{equation}
The covariance matrix $\sigma=V-\mathbf{dd}^\intercal$ then satisfies
\begin{equation}
    \dot{\sigma} = K\sigma+\sigma K^\intercal + \mathcal{D}
    \qquad\text{which is solved by}\qquad
    \sigma(t) = S(t)\sigma(0)S(t)^\intercal + \int_0^t \dd s\, S(t-s) \mathcal{D} S(t-s)^\intercal .
 \end{equation}
Here we have defined
\begin{equation}
    S(t) = e^{Kt}
    \qquad\text{and}\qquad
    \mathbf{h}(t)=\int_0^t\dd s\, S(t-s)\Omega\mathbf{g} .
\end{equation}

\subsection{Stochastic noise in the trapping}\label{apx:gauss-trap-variations}
Using that the noise has vanishing mean, i.e. $\mean{\xi_i(t)} = 0$, we find that the dynamics are given by
\begin{align}
    \mean{\mathbf{d}(t)}_\xi &= S(t) \mathbf{d}(0) \\
    \mean{V(t)}_\xi &= S(t) V(0) S(t)^\intercal + \mean{\mathbf{h}(t)\mathbf{h}(t)}_\xi .
\end{align}
The covariance $\Cov_\xi[\mathbf{d}]=\mean{\mathbf{dd}^\intercal}_\xi - \mean{\mathbf{d}}_\xi\mean{\mathbf{d}}_\xi^\intercal$ is given by
\begin{equation}\label{eq:apx:trap-variations-covariance}
    \Cov_\xi[\mathbf{d}] = \int_0^t\dd s \int_0^t\dd s' \, S(t-s) B \, C(s-s') \, B^\intercal S(t-s')^\intercal 
\end{equation}
where $B$ includes the linear interaction and is defined by $\mathbf{g}=B\,\xi$. 
For the Hamiltonian given by \eqref{eq:trap-complete-hamiltonian}, it is given by
\begin{equation}
    B=\begin{pmatrix}
        0  & 0 \\ 1 & 0 \\ 0 & 0 \\ 0 & 1
    \end{pmatrix} 
    \quad\quad\text{with correlator}\quad\quad
    C(\tau) =\begin{pmatrix}
        C_A(\tau) & 0 \\
        0 & C_B(\tau)
    \end{pmatrix}
\end{equation}

\subsubsection{Quasi-Static limit ($t_c \gg t$)}
The correlator is approximately constant $C_i(\tau)=\Delta \xi^2$ over an entire single-run dynamics.
Only classical variations between runs induce noise and $\xi_i(t)=\mathrm{const.}$ over the entire run.
\Eqref{eq:apx:trap-variations-covariance} becomes
\begin{equation}
    \Cov_\xi[\mathbf{d}]=\Delta\xi^2 \sum_{i=A,B}\left(\int_0^t\dd s\, S(t-s)B_i\right)\left(\int_0^t\dd s'\, S(t-s)B_i\right)^\intercal
\end{equation}
with $B_A=(0,1,0,0)^\intercal$ and $B_B=(0,0,0,1)^\intercal$.

\subsubsection{Markovian limit ($t_c \rightarrow 0$)}
By taking $C_i(\tau)=2D_i\delta(\tau)$, we find 
\begin{equation}
    \Cov_\xi[\mathbf{d}] = \int_0^t\dd s\, S(t-s) \mathcal{D} S(t-s)^\intercal
    \quad\text{with}\quad
    \mathcal{D}=\bigoplus_i\begin{pmatrix}
        0 & 0 \\
        0 & 2D_i
    \end{pmatrix} .
\end{equation}
Differentiating \Eqref{eq:mean-covariance-matrix}, we find the following Markovian master equation governing the evolution
\begin{equation}
    \dv{\sigma}{t} = K\sigma + \sigma K^\intercal + \mathcal{D} .
\end{equation}

This is equivalent to the Lindblad equation
\begin{equation}
    \dv{\rho}{t} = -\frac{i}{\hbar}[\op H,\rho] - \sum_i\frac{D_i}{\hbar^2}[\op x_i,[\op x_i,\rho]].
\end{equation}
The Lindblad operators of this can be rewritten like
\begin{equation}
    L_i = \sqrt{\frac{2D_i}{\hbar^2}}x_i = \mathbf{l}_i^\intercal\mathbf{r},\quad \mathbf{l}_i = \sqrt{2D_i/\hbar^2}\, \mathbf{e}_{x_i},
\end{equation}
where $\mathbf{e}_{x_i}$ picks out the $x_i$ component.
The Kossakowski matrix $C=\sum_i \mathbf{l}_i\mathbf{l}_i^\intercal$ is purely real and the diffusion matrix $\mathcal{D}$ is given by
\begin{equation}
    \mathcal{D} = \hbar^2\Omega \left(\sum_i\frac{2D_i}{\hbar^2}\mathbf{e}_{x_i}\mathbf{e}_{x_i}^\intercal\right)\Omega^\intercal = 2 \sum_i D_i \mathbf{e}_{p_i}\mathbf{e}_{p_i}^\intercal = 2\bigoplus_i\begin{pmatrix}
        0&0\\0&D_i
    \end{pmatrix} .
\end{equation}

\subsection{Entanglement dynamics for Gaussian states}\label{apx:entanglement-gaussian}
Starting from the quadratically expanded gravitational Hamiltonian \Eqref{eq:apx:gravity-hamiltonian} with $\theta_A=\theta_B\equiv\theta$:
\begin{equation}
    \op H_\mathrm{Gravity} \approx \lambda\left[
    \left(\sin^2\theta-\frac{1}{2}\cos^2\theta\right)\op x_A \op x_B 
    + \left(\cos^2\theta-\frac{1}{2}\sin^2\theta\right)\op y_A \op y_B
    - \left(\frac{3}{2}\cos\theta\sin\theta\right)\left(\op x_A \op y_B + \op x_B \op y_A\right)
    \right] .
\end{equation}
To simplify the calculation, we only take terms coupling both particles $A$ and $B$, as all other terms cannot influence the bipartite entanglement generation.
A more sophisticated analysis and numerical data show, that in fact, this is true.
We can construct the Gaussianity-preserving Hamiltonian $\op H=\tfrac{1}{2}\mathbf{r}^\intercal G \mathbf{r}$, where $G$ is given in block-matrix form as
\begin{equation}
    G = \begin{pmatrix}
        0 & \tilde{G} \\
        \tilde{G} & 0
    \end{pmatrix}
    \quad\text{where}\quad
    \tilde{G} = \begin{pmatrix}
        A & 0 & C & 0\\
        0&0&0&0\\
        C & 0 & B & 0\\
        0&0&0&0
    \end{pmatrix} .
\end{equation}
Here, we denote $A=\lambda (\sin^2\theta-\tfrac{1}{2}\cos^2\theta)$, $B=\lambda(\cos^2\theta - \tfrac{1}{2}\sin^2\theta)$ and $C = -\tfrac{3\lambda}{2}\sin\theta\cos\theta$.
Computing $\sigma(t)=e^{Kt}\sigma_0e^{K^\intercal t}$ with $\sigma_0$ given by \Eqref{eq:gaussian-initial-state-d} and \Eqref{eq:gaussian-initial-state-V} and partially transposing the covariance matrix to $\sigma(t)^\Gamma = P\sigma(t)P$ with $P=\diag\{1,1,1,1,1,-1,1,-1\}$, we are left with a block-matrix in the form of
\begin{equation}
    \sigma(t)^\Gamma = \begin{pmatrix}
       \sigma_1 & \sigma_2 \\
       -\sigma_2 & \sigma_1
    \end{pmatrix},
\end{equation}
whose eigenvalues are given by $\{\lambda_i,\lambda_i^*\}$, when $\lambda_i$ is an eigenvalue of the $4\times4$ matrix $\sigma_1 + i\sigma_2$.
The symplectic eigenvalues of $\sigma(t)^\Gamma$ are (in linear order in $t$) given by
\begin{equation}
    \left\{\abs{A t \Delta x_A\Delta x_B (1+2\bar{n}_x) \pm \frac{1}{2}(1+2\bar{n}_x)}, \abs{B t \Delta y_A\Delta y_B \pm \frac{1}{2}}\right\}
\end{equation}
with multiplicity 2 for each eigenvalue.
This is independent of the parameter $C$ for all $\theta$.
The logarithmic negativity can be computed with \Eqref{eq:log-neg-gaussian} and is given in linear order in $t$ at zero temperature by
\begin{equation}
    E_N = \frac{2 t}{\hbar\log2}\big(\abs{A} \Delta x_A\Delta x_B + \abs{B} \Delta y_A\Delta y_B\big) .
\end{equation}

\subsection{Corrections to non-Gaussianity}\label{apx:non-gaussianity}
After classical averaging, the resulting state $\mean{\rho} \equiv \rho$ is in general no longer Gaussian.
In lowest order, the deviation from the Gaussian reference state $\rho_G$ can be written as
\begin{equation}
    \rho = \rho_G + \Delta\rho 
    \quad \text{with} \quad 
    \Delta \rho = \int\dd\xi \, p(\xi) \left(D(\xi)\rho_G D^\dagger(\xi) - \rho_G\right) .
\end{equation}
Using the reverse triangle inequality for the trace norm,
\begin{equation}
    \norm{\rho}_1 \geq \norm{\rho_G}_1 - \norm{\Delta \rho}_1 .
\end{equation}
and applying the monotonicity of the logarithm, we obtain the following lower bound on the logarithmic negativity:
\begin{equation}
    E_N(\rho)=\log_2\norm{\rho^\Gamma}_1 \geq \log_2(\norm{\rho_G^\Gamma}_1 - \norm{\Delta\rho^\Gamma}_1) = \log_2\norm{\rho_G^\Gamma}_1 - \frac{\norm{\Delta\rho^\Gamma}_1}{\log2 \norm{\rho_G^\Gamma}_1} + \mathcal{O}(\norm{\Delta\rho}_1^2) .
\end{equation}
Here, we have expanded the logarithm to first order for small deviations $\norm{\Delta\rho}_1 \ll \norm{\rho_G}_1$.
The random displacement channel given by $D(\xi)$ is LOCC and the logarithmic negativity is non-increasing under LOCC channels as shown in Ref.~\cite{Plenio2005}, thus $E_N(\rho)\leq E_N(\rho_G)$.
Combining both bounds, the deviation from the Gaussian logarithmic negativity is bounded by
\begin{equation}
    \Delta E_N = E_N(\rho_G) - E_N(\rho) \leq \frac{\norm{\Delta \rho^\Gamma}_1}{\log2\norm{\rho_G^\Gamma}_1} = \frac{\norm{\Delta\rho^\Gamma}_1}{\log 2} 2^{-E_N(\rho_G)} + \mathcal{O}(\norm{\Delta\rho}_1^2).
\end{equation}

To estimate the magnitude of the non-Gaussian correction $\Delta\rho$, we express it in terms of the characteristic function of the quadrature operators $\op x$ and $\op p$ by
\begin{equation}\label{eq:apx:delta-rho}
    \Delta \rho = \frac{1}{(2\pi)^n} \int \dd^{2n}\lambda \, \Delta\chi(\lambda) e^{-i\lambda^\intercal\Omega \mathbf{r}},
\end{equation}
where $\Delta \chi(\lambda) = \chi(\lambda)-\chi_G(\lambda)$ denotes the difference between the characteristic function $\chi(\lambda)=\int\dd\xi\,p(\xi)\chi_\xi(\lambda)$ and and its Gaussian counterpart $\chi_G(\lambda)$.
The characteristic function can be written as a cumulant expansion
\begin{equation}
    \chi(\lambda) = \mean{e^{f(\xi;\lambda)}}_\xi = \exp{\sum_{n=0}^\infty \frac{\kappa^{(n)}(f)}{n!}} 
    \quad\text{with}\quad 
    f(\xi;\lambda)=-\frac{1}{2}\lambda^\intercal\Omega^\intercal \sigma(\xi) \Omega\lambda + i\lambda^\intercal \Omega d(\xi) .
\end{equation}
Expanding $f = f_0 + f_1 \xi + f_2 \xi^2 + \mathcal{O}(\xi^3)$ in powers of the classical variable $\xi$ and assuming a Gaussian distribution $\xi\sim\mathcal{N}(0,\Delta\xi^2)$, Wick's theorem yields $\mean{\xi^{2m+1}}=0$ and $\mean{\xi^{2m}}=(2m-1)!!(\Delta \xi^2)^m$.
Consequently, the cumulants scale as $\kappa^{(n)}\sim\mathcal{O}(\Delta\xi^{2\lceil n/2\rceil})$ where the lowest order is explicitly given by
\begin{equation}
    \chi(\lambda) = \chi_G(\lambda) \exp{\frac{\Delta \xi^2}{2}(2f_1^2 + f_2) + \mathcal{O}(\Delta \xi^4)} \approx \chi_G(\lambda) \left(1+\frac{\Delta \xi^2}{2}(2f_1^2 + f_2)\right) + \mathcal{O}(\Delta \xi^4) .
\end{equation}
Thus, the non-Gaussian correction $\norm{\Delta \rho}_1 \sim \Delta \xi^2 \norm{\rho_G}_1$, with a pre-factor determined by $2f_1(\lambda)^2 + f_2(\lambda)$ and the integral expression from \Eqref{eq:apx:delta-rho}.
For small variations, the state is therefore almost Gaussian with a small correction to the logarithmic negativity
\begin{equation}
    \Delta E_N \sim \Delta \xi^2 + \mathcal{O}(\Delta \xi^4) .
\end{equation}
\section{Cat-states}\label{apx:cat-states}
\subsection{Entanglement dynamics for two-level cat-states}\label{apx:cat-states-gravity}
Starting from the Hamiltonian \Eqref{eq:apx:gravity-hamiltonian}, we take the limit $\op{y} \rightarrow 0$ and expand to second order in $\op{x}$ to find
\begin{multline}\label{eq:apx:2-level-cat-Hamiltonian}
    \op H_\mathrm{Gravity} \approx \frac{GM_AM_B}{2L} + \lambda \bigg(
    \op x_A \op x_B \sin\theta_A\sin\theta_B  
    - \frac{1}{2}\op x_A \op x_B \cos\theta_A\cos\theta_B \\
    - \frac{1}{2}\left(\op x_A^2 + \op x_B^2\right) 
    + \frac{3}{4} \left(\op x_A^2 \cos^2\theta_A + \op x_B^2 \cos^2\theta_B \right)
    + L\op x_A \sin\theta_A - L\op x_B \sin\theta_B
    \bigg).
\end{multline}
Upon discretizing the position operator by $\op{x}\rightarrow \Delta x \sigma_z$ with $\sigma_z$ being the Pauli matrix, the interaction reduces to an effective two-level Hamiltonian acting on the Hilbert-space spanned by $\left\{\ket{\psi^{(1)}},\ket{\psi^{(2)}}\right\}^{\otimes 2}$.
For the initial product state $\rho_\mathrm{Particles}=\frac{1}{4}(\identity+\sigma_x)^{\otimes 2}$, the evolution under $\op{H}_\mathrm{Gravity}$ takes the analytic form
\begin{equation}\label{eq:apx:two-level-no-casimir}
\rho_\mathrm{Gravity}(t) =\frac{1}{4} \begin{pmatrix}
    1 & e^{-i t (A - B_B)/\hbar} & e^{-i t (A+B_A)/\hbar} & e^{-i t (B_A-B_B)/\hbar}\\
    e^{i t (A-B_B)/\hbar} & 1 & e^{-i t (B_A+B_B)/\hbar} & e^{i t (A-B_A)/\hbar}\\
    e^{i t (A+B_A)/\hbar} & e^{i t (B_A+B_B)/\hbar} & 1 & e^{i t (A+B_B)/\hbar}\\
    e^{i t (B_A-B_B)/\hbar} & e^{-i t (A-B_A)/\hbar} & e^{-i t (A+B_B)/\hbar} & 1
\end{pmatrix},
\end{equation}
with $A=2\lambda\Delta x_A\Delta x_B\left(\sin\theta_A\sin\theta_B-\tfrac{1}{2}\cos\theta_A\cos\theta_B\right)$ and $B_i = 2 L \lambda \Delta x_i \sin\theta_i$.
The eigenvalues of the partial transposed state $\rho^\Gamma$ are given by
\begin{equation}
    \left\{\frac{1}{2}\sin(\frac{t A}{\hbar}),-\frac{1}{2}\sin(\frac{t A}{\hbar}),\sin^2\left(\frac{t A}{2\hbar}\right),\cos^2\left(\frac{t A}{2\hbar}\right)\right\},
\end{equation}
thus resulting in a logarithmic negativity of
\begin{equation}
    E_N = \log_2\left(1+\abs{\sin(\phi t)}\right)
\end{equation}
with 
\begin{equation}
    \phi = \frac{A}{\hbar} = \frac{\lambda \Delta x_A\Delta x_B}{\hbar}\left(2\sin\theta_A\sin\theta_B - \cos\theta_A\cos\theta_B\right)\, .
\end{equation}

\subsection{Stochastic placement variations for two-level cat-states}\label{apx:cat-states-variations}
In addition to the gravitational interaction \Eqref{eq:apx:2-level-cat-Hamiltonian}, we include Casimir and magnetic-dipole interactions from \Eqref{eq:apx:casimir-hamiltonian} and \Eqref{eq:apx:mag-dipole-hamiltonian} at arbitrary orientations.
In the limit $\op{y}\rightarrow 0$, we can expand to \textit{first order} in the variations and find
\begin{multline}
    \op{H} = \op{H}_\mathrm{Gravity} +
    \eta\left[6 \op{x}_A  \xi_A^{(L)} \sin\theta_A - 2(L-R-\tfrac{d_s}{2}) \op{x}_A \xi_A^{(\theta)} \cos\theta_A + \frac{12 \op{x}_A^2 \xi_A^{(L)} \sin^2\theta_A}{L-R-d_s/2} - 6 \op{x}_A^2 \xi_A^{(\theta)} \sin\theta_A\cos\theta_A \right] \\
    + \eta\left[-6 \op{x}_B  \xi_B^{(L)} \sin\theta_B + 2(L-R-\tfrac{d_s}{2})\op{x}_B \xi_B^{(\theta)} \cos\theta_B + \frac{12 \op{x}_B^2 \xi_B^{(L)} \sin^2\theta_B}{L-R-d_s/2} - 6 \op{x}_B^2 \xi_B^{(\theta)} \sin\theta_B\cos\theta_B \right]
\end{multline}
for Casimir interactions and
\begin{multline}
    \op H = \op{H}_\mathrm{Gravity} 
    + \delta\left[12 \op{x}_A  \xi_A^{(L)} \sin\theta_A - 3(L-\tfrac{d_s}{2}) \op{x}_A \xi_A^{(\theta)} \cos\theta_A + \frac{30 \op{x}_A^2 \xi_A^{(L)} \sin^2\theta_A}{L-d_s/2} - 12 \op{x}_A^2 \xi_A^{(\theta)} \sin\theta_A\cos\theta_A \right] \\
    + \delta\left[-12 \op{x}_B  \xi_B^{(L)} \sin\theta_B + 3(L-\tfrac{d_s}{2}) \op{x}_B \xi_B^{(\theta)} \cos\theta_B + \frac{30 \op{x}_B^2 \xi_B^{(L)} \sin^2\theta_B}{L-d_s/2} - 12 \op{x}_B^2 \xi_B^{(\theta)} \sin\theta_B\cos\theta_B \right]
\end{multline}
for magnetic-dipole interactions.

After replacing $\op{x}\rightarrow \Delta x \sigma_z$ to project into the two-dimensional Hilbert-space for the two-level cat-states, the time evolved and averaged state can be obtained analytically.
Here, we again distinguish between the general case, where the detectors move from run to run and the case, where the shield moves and thus $\xi^{(L)}_A=-\xi^{(L)}_B$ and $\xi^{(\theta)}_A=\xi^{(\theta)}_B$:

\subsubsection{Detector movement}
In the general case with uncorrelated $\xi_{A/B}$, the averaged density matrix can be written in the form
\begin{equation}
    \mean{\rho} = \rho_\mathrm{Gravity}(t) \odot \begin{pmatrix}
        1 & e^{-\sum_k \gamma_{B,k}^2} & e^{-\sum_k\gamma_{A,k}^2} & e^{-\sum_k(\gamma_{A,k}^2+\gamma_{B,k}^2)} \\
        e^{-\sum_k\gamma_{B,k}^2} & 1 & e^{-\sum_k(\gamma_{A,k}^2+\gamma_{B,k}^2)} & e^{-\sum_k \gamma_{A,k}^2} \\
        e^{-\sum_k\gamma_{A,k}^2} & e^{-\sum_k (\gamma_{A,k}^2+\gamma_{B,k}^2)} & 1 & e^{-\sum_k \gamma_{B,k}^2} \\
        e^{-\sum_k(\gamma_{A,k}^2+\gamma_{B,k}^2)} & e^{-\sum_k \gamma_{A,k}^2} & e^{-\sum_k \gamma_{B,k}^2} & 1
    \end{pmatrix}
\end{equation}
where $k\in \{L,\theta\}\times\{\mathrm{Casimir},\ \mathrm{mag.\,Dipole}\}$ iterates over all considered interactions and variations with
\begin{align}
    \gamma_{i,\,\theta,\,\mathrm{Casimir}} &= \sqrt{8}(L-R-\frac{d_s}{2})\eta\Delta\theta\cos\theta_i\, \Delta x_i\, t\,/\,\hbar,
     \quad \quad \gamma_{i,\,L,\,\mathrm{Casimir}} = \pm\sqrt{72} \eta \Delta L \sin\theta_i\, \Delta x_i\, t\,/\,\hbar\\
    \gamma_{i,\,\theta,\,\mathrm{mag.\,Dipole}} &= \sqrt{18}(L-\frac{d_s}{2})\delta\Delta\theta_i\cos\theta_i\, \Delta x_i\, t\,/\,\hbar,
    \quad \quad \gamma_{i,\,L,\,\mathrm{mag.\,Dipole}} = \pm\sqrt{288} \delta \Delta L \sin\theta_i\, \Delta x_i\, t\,/\,\hbar
\end{align}
The $\pm$ distinguishes between particles $A$ and $B$ and $\odot$ denotes the element-wise Hadamard product of the two matrices, where $\rho_\mathrm{Gravity}(t)$ is given by \Eqref{eq:apx:two-level-no-casimir} and is the evolved state with only gravitational interactions included.
Denoting $\gamma_i^2 \equiv \sum_k \gamma_{i,k}^2$, the logarithmic negativity of this state is given by
\begin{equation}
    E_N = \max\left\{0, \log_2\left[\frac{1}{2}\left(1+e^{-\gamma_A^2-\gamma_B^2}+\sqrt{e^{-2\gamma_A^2}+e^{-2\gamma_B^2}-2e^{-\gamma_A^2-\gamma_B^2}\cos(2\phi t)}\right)\right]\right\}\, .
\end{equation}
In the special case of $\gamma_A = \gamma_B := \gamma$, corresponding to two identical delocalizations, the logarithmic negativity can be expressed in the form
\begin{align}
    E_N &= \max\left\{0, \log_2\left(e^{-\gamma^2} (\cosh(\gamma^2) + \abs{\sin\phi t})\right)\right\} \\
    &= \log_2\left\{\frac{1}{2}e^{-\gamma^2}\left(\abs{\sin\phi t - \sinh(\gamma^2)} + \abs{\sin\phi t + \sinh(\gamma^2)} + 2\cosh(\gamma^2)\right)\right\}\, .
\end{align}

\subsubsection{Shield movement}
If $\xi_{A/B}$ are correlated due to the shield movement effecting the couplings of both particles with the shield in the same way, the averaged density matrix is given by
\begin{equation}
    \mean{\rho} = \rho_\mathrm{Gravity}(t) \odot \begin{pmatrix}
        1 & e^{-\sum_k \gamma_{B,k}^2} & e^{-\sum_k \gamma_{A,k}^2} & e^{-\sum_k(\gamma_{A,k}-\gamma_{B,k})^2} \\
        e^{-\sum_k \gamma_{B,k}^2} & 1 & e^{-\sum_k(\gamma_{A,k}+\gamma_{B,k})^2} & e^{-\sum_k \gamma_{A,k}^2} \\
        e^{-\sum_k \gamma_{A,k}^2} & e^{-\sum_k(\gamma_{A,k}+\gamma_{B,k})^2} & 1 & e^{-\sum_k \gamma_{B,k}^2} \\
        e^{-\sum_k(\gamma_{A,k}-\gamma_{B,k})^2} & e^{-\sum_k \gamma_{A,k}^2} & e^{-\sum_k \gamma_{B,k}^2} & 1
    \end{pmatrix}
\end{equation}
with $\gamma_{i,\,k}$ given from before.
A compact expression for the logarithmic negativity of the state cannot be given here.
For only one considered interaction, however, it is given by
\begin{multline}
    E_N = \max\bigg\{0,\log_2\bigg[\frac{1}{2}\Big(1+e^{-\gamma_A^2-\gamma_B^2}\cosh(2\gamma_A\gamma_B) + \\
    \sqrt{e^{-2(\gamma_A^2+\gamma_B^2)}\sinh(2\gamma_A\gamma_B) + e^{-2\gamma_A^2} + e^{-2\gamma_B^2} - 2e^{-\gamma_A^2-\gamma_B^2}\cos(2\phi t)}\Big)\bigg]\bigg\}
\end{multline}
In the special case $\gamma_A = \gamma_B := \gamma$, this reduces to
\begin{equation}
    E_N = \max\left\{0,\log_2\left[\frac{1}{4}\left(3+e^{-4\gamma^2}+2e^{-2\gamma^2}\sqrt{4e^{2\gamma^2}\sin^2(\phi t) + \sinh^2(2\gamma^2)}\right)\right]\right\}
\end{equation}

\subsection{Numerical results for large superpositions}
\begin{figure*}[!th]
    \centering
    \includegraphics[width=\linewidth]{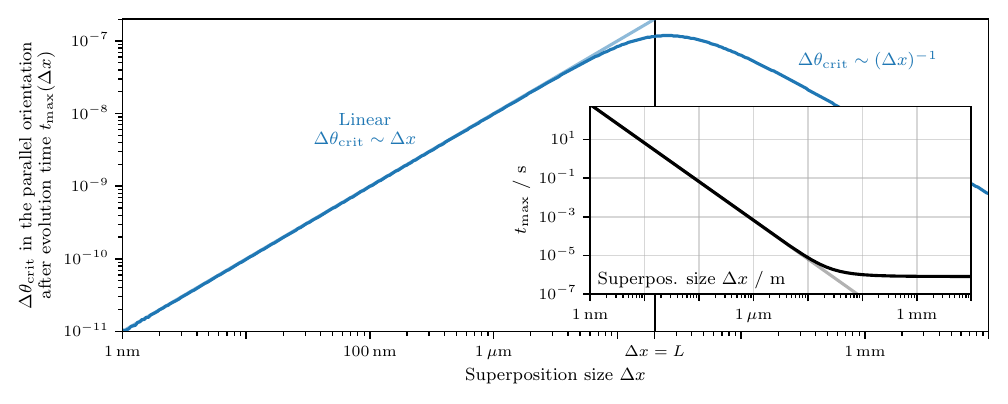}
    \caption{\justifying\small 
    \textbf{Stability for large spatial superpositions.}
    Numerically determined critical angular variations $\Delta \theta_\mathrm{crit}$ as a function of the superposition size $\Delta x$, evaluated at the interaction time $t_\mathrm{max}(\Delta x)$ at which the entanglement is maximal.
    For $\Delta \theta > \Delta \theta_\mathrm{crit}$ the logarithmic negativity vanishes.
    The inset shows the corresponding times $t_\mathrm{max}$.
    A qualitative change in stability occurs for $\Delta x \gtrsim L$, where decoherence starts to dominate.
    This analysis was exemplarily performed for silica particles prepared in idealized two-level cat-states.
    }
    \label{fig:apx:large-delta-x}
\end{figure*}
Throughout this work, we always employ the approximation $\Delta x \ll L$, which allows for the interaction potentials to be expanded up to second order in the position operators (see e.g. Apx.~\ref{apx:interactions}).
For superposition sizes approaching the micrometer scale, however, this approximation breaks down and the analytical treatment is no longer valid.
To assess the robustness of entanglement generation beyond this regime, we therefore perform a full numerical analysis for arbitrary superposition sizes $\Delta x$.
\figref{fig:apx:large-delta-x} shows the numerically determined critical angular standard deviation $\Delta \theta_\mathrm{crit}$ as a function of $\Delta x$.
For angular fluctuations exceeding $\Delta \theta > \Delta \theta_\mathrm{crit}$, the logarithmic negativity vanishes and no bipartite entanglement is present in the averaged state.
The entire analysis is exemplarily performed for silica particles prepared in idealized two-level cat-states interacting with the shield via Casimir interactions.
The data is evaluated at an evolution time $t_\mathrm{max}(\Delta x)$, defined as the time at which the entanglement reaches its maximal value of $E_N(t_\mathrm{max})=1$ in the absence of noise.
This time is shown in the inset of \figref{fig:apx:large-delta-x} as a function of $\Delta x$.

For small superposition sizes $\Delta x \ll L$, this time scales as $t_\mathrm{max}\sim \Delta x^{-2}$, in agreement with the analytical results in \Eqref{eq:cat-state-entanglement}.
From \Eqref{eq:log-neg-cat-variations}, we find that at critical variations $\gamma_\mathrm{crit}(t_\mathrm{max})$, it has to hold that $\Delta \theta_\mathrm{crit} \propto 1/(\Delta x\, t_\mathrm{max})$, such that the reductioTn in the interaction time overcompensates the increased sensitivity.
As a result, the critical angular tolerance \textit{increases} linearly with $\Delta x$, leading to more stability.
This behavior changes qualitatively once the superposition site exceeds the particle separation $\Delta x \gtrsim L$.
Here, the time $t_\mathrm{max}$ saturates and no longer decreases with $\Delta x$.
Consequently, decoherence continues to grow with $\Delta x$, resulting in a loss of stability.

While increasing the spatially delocalization can, within our model, enhance robustness against placement-induced decoherence, large superpositions are nevertheless known to be fragile under other decoherence mechanisms.
In particular, Ref.~\cite{Kerker2020} shows that Coulomb-induced decoherence from nearby conducting surfaces increases rapidly with growing $\Delta x$.
This limitation is rooted in a different physical mechanism than the placement-induced dephasing considered here and therefore, one might not be able to fully exploit the parametric regime in which larger $\Delta x$ would otherwise improve robustness.
More generally, for scattering processes such as collisions with air molecules or photons, the decoherence rate typically to increase with $\Delta x^2$~\cite{Schlosshauer2010}.

\subsection{Cat-states in front of a vibrating thermal shield}\label{apx:cat-states-shield}
The Hamiltonian 
\begin{equation}
    \op{H}_\mathrm{Shield} = \sum_{(k,l)}\op{H}_{kl} + \hbar\omega_{kl}(\op{a}^\dagger_{kl}\op{a}_{kl}+1/2)
\end{equation} 
describes the vibrating shield modes labeled by $(k,l)$ and their interaction with the particles.
As gravity couples only through position operators, $\hat H_\mathrm{Gravity}$ commutes with $\hat H_\mathrm{Shield}$.
As the time evolution generated by $\hat H_\mathrm{Gravity}$ has already been analyzed in Appendix~\ref{apx:cat-states-gravity}, it remains only to determine the evolution under $\hat H_\mathrm{Shield}$.
In the interaction picture w.r.t. $\op{H}_0=\sum_{(k,l)}\hbar\omega_{kl}(\op{a}^\dagger_{kl}\op{a}_{kl} + 1/2)$ it is given by
\begin{equation}
    \op{H}_\mathrm{Shield,int} = \sum_{(k,l)} \op{G}_{kl} (\op{a}_{kl}e^{-i\omega_{kl}t} + \op{a}_{kl}^\dagger e^{i\omega_{kl}t}) ,
\end{equation}
where $\hat G_\mathrm{kl}=\diag(g_1,\dots,g_4)$ incorporates the interaction of the delocalized particles with the shield.
It can be derived by making the replacements $\op{x}\rightarrow\Delta x \sigma_z$ and $\op{y} \rightarrow 0$ in \Eqref{eq:hamiltonian-particle-shield-parallel} and \Eqref{eq:hamiltonian-particle-shield-linear} and it thus given by
\begin{align}
    \op{G}_{kl, \mathrm{Casimir,\,parallel}} &= 4 \eta \Delta x (L-R-\tfrac{d_s}{2}) \, \partial_r u_{kl}\big|_{r=0}\sqrt{\frac{\hbar}{2m_\mathrm{eff}\omega_{kl}}}\,\diag(0, -1, 1, 0)\, , \\
    \op{G}_{kl, \mathrm{Casimir\,linear}} &= 12 \eta \Delta x \, u_{kl}(r=0)\sqrt{\frac{\hbar}{2m_\mathrm{eff}\omega_{kl}}}\,\diag(1, 0, 0, -1)
\end{align}
for Casimir interactions and by
\begin{align}
    \op{G}_{kl, \mathrm{mag.\,dipole,\ parall.}} &= 6 \delta \Delta x (L-\tfrac{d_s}{2}) \, \partial_r u_{kl}\big|_{r=0}\sqrt{\frac{\hbar}{2m_\mathrm{eff}\omega_{kl}}}\,\diag(0, -1, 1, 0)\, , \\
    \op{G}_{kl, \mathrm{mag.\,dipole,\ linear}} &= 24 \delta \Delta x \, u_{kl}(r=0)\sqrt{\frac{\hbar}{2m_\mathrm{eff}\omega_{kl}}}\,\diag(1, 0, 0, -1)
\end{align}
for magnetic-dipole interactions.

As the Magnus expansion breaks off in second order, the time evolution can be found exactly and is given by
\begin{align}
    \op{U}(t) &= \exp{-\frac{i}{\hbar}\int_0^t\dd t_1\,\op{H}(t_1) - \frac{1}{2\hbar^2}\int_0^t\dd t_1 \int_0^{t_1}\dd t_2 \, [\op{H}(t_1),\op{H}(t_2)]} \\
    &= \exp{\op{G}(f_1 \op{a}^\dagger - f_1^* \op{a}) + i\op{G}^2f_2} \\
    &= \diag\left(\op{D}(f_1 g_1)e^{if_2g_1^2},\op{D}(f_1 g_2)e^{if_2g_2^2},\op{D}(f_1 g_3)e^{if_2g_3^2},\op{D}(f_1 g_4)e^{if_2g_4^2}\right)
\end{align}
with 
\begin{equation}
    f_1 = \frac{(1-e^{i\omega_{kl}t})}{\hbar \omega_{kl}} \quad \text{and} \quad f_2=\frac{t\omega_{kl} - \sin(t\omega_{kl})}{\hbar^2\omega_{kl}^2}.
\end{equation}
The displacement operator is defined by $\op{D}(\alpha)=\exp{\alpha \op{a}^\dagger-\alpha^*\op{a}}$. The time evolved density matrix $\rho(t)=\op{U}(t)\rho_0\op{U}^\dagger(t)$ of $\rho_0=\bigotimes_{(k,l)}\rho_{\mathrm{th},kl}\otimes\rho_\mathrm{Particles}$ is therefore given by
\begin{equation}
    \rho_{ij}(t) = \left[\bigotimes_{(k,l)}\op{D}(f_1 g_i) \rho_{\mathrm{th},kl}\op{D}^\dagger(f_1 g_j)\right] e^{if_2 (g_i^2 - g_j^2)}\rho_{\mathrm{particles},ij}
\end{equation}
where the indices $i,j$ of the resulting block-matrix $\rho(t)$ label the basis states of the particles. 
Tracing out the shield and using $\tr{A\otimes B}=\tr A \tr B$ one finds for the time evolution of both particles
\begin{equation}
    \rho_{\mathrm{particles},ij}(t) = \prod_{(k,l)}\tr{\op{D}(f_1g_i)\rho_{\mathrm{th},kl}\op{D}^\dagger(f_1g_j)}e^{if_2(g_i^2 - g_j^2)}\rho_{\mathrm{Particles},ij} .
\end{equation}
Using that $\tr{\op{D}(\zeta_i)\rho_\mathrm{th}\op{D}^\dagger(\zeta_j)} = \exp{\tilde{\zeta} - \abs{\Delta \zeta}^2\left(\frac{1}{2}+\bar{n}\right)}$ where $\Delta \zeta = \zeta_i - \zeta_j$ and $\tilde{\zeta}=(\zeta_j^*\zeta_i-\zeta_j\zeta_i^*)/2 = 0$~\cite{Steiner2025}, the elements of the evolved density matrix are finally given by
\begin{equation}
    \rho_{\mathrm{particles},ij}(t) = \exp\left\{-\sum_{(k,l)}\abs{g_{i,kl} - g_{j,kl}}^2 f_1f_1^* \left(\frac{1}{2}+\bar{n}\right)\right\}e^{if_2(g_{i,kl}^2 - g_{j,kl}^2)}\rho_{\mathrm{particles},ij}(0)\, .
\end{equation}
Crucially, ground-state fluctuations of the shield give rise to an additional effect even at $T=0$.

\begin{figure}[!th]
    \centering
    \includegraphics[width=\linewidth]{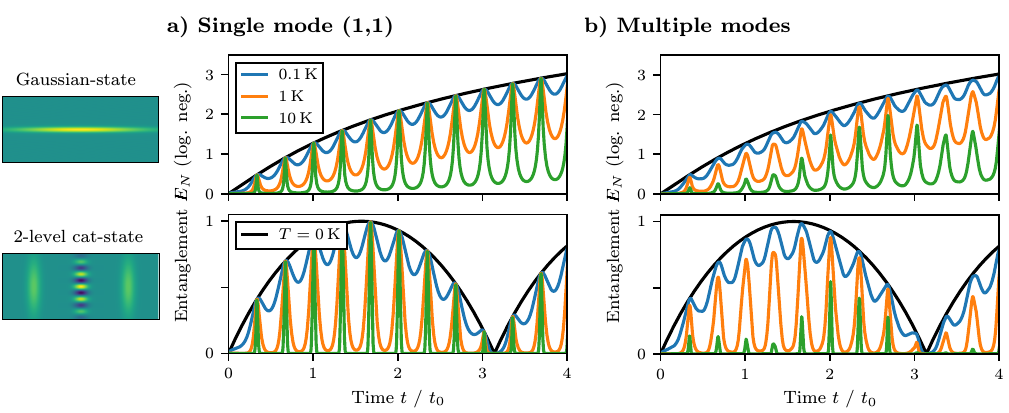}
    \caption{\justifying\small 
    \textbf{Entanglement dynamics in a GIE setup in the presence of a quantum shield.}
    Results are shown for delocalized silica particles in the parallel orientation $(\theta = 0)$.
    This figure is quantitatively the same as \figref{fig:shield-parallel-magnetic-dipole} but with weaker Casimir interactions between the particles and the shield.
    Here, the particle parameters from \tabref{tab:parameters} were used with a Copper shield of size $d_s=500\si{nm}$ and $r_s = 1\si{cm}$.
    The thinner thickness compared to \figref{fig:shield-parallel-magnetic-dipole} was chosen to better visualize the underlying dynamics for the much weaker gravitational coupling between the silica particles.
    \textbf{a,} Entanglement dynamics for Gaussian states (\textbf{top}) and two-level cat-states (\textbf{bottom}) with the shield dynamics restricted to the dominant $(1,1)$-mode only.
    \textbf{b,} Corresponding dynamics where multiple shield modes are included (here: lowest 64 modes).
    }
    \label{fig:apx:shield-parallel-casimir}
\end{figure}

\end{document}